%% file: esa2014.tex
\documentclass{llncs}

\input{preamble}

\title{Bank Conflict Free Comparison-based Sorting On GPUs}

\author{Nodari Sitchinava \inst{1}
   \and
   Volker Weichert \inst{2}
   }

\institute{
   University of Hawaii, Manoa, HI, USA
\and
   Goethe University Frankfurt am Main, Frankfurt, Germany
}
\date{}

\begin{document}
\maketitle
\input{abstract}





\input{intro}


\input{framework}

\input{sorting}

\input{merging}

\input{experiments}

\input{conclusion}



\bibliographystyle{splncs03}
\bibliography{refs}


\newpage

\begin{appendix}

\input{prefix_sums}
\input{proof-lemma-2}
\newpage
\input{additional-charts}
\input{register-optimization}
\input{shearsort}
\input{pseudocode}

\end{appendix}

\end{document}

%% file: preamble.tex
\usepackage{amsfonts}
\usepackage{amsmath}
\usepackage{url}
\usepackage{graphicx}
\usepackage{ifthen}
\usepackage{algorithm}
\usepackage{algorithmic}

\usepackage{tikz}
\usetikzlibrary{shapes,arrows,decorations,decorations.pathmorphing,decorations.pathreplacing,plotmarks}

\usepackage{subfig}

\newcommand\blfootnote[1]{%
  \begingroup
  \renewcommand\thefootnote{}\footnote{#1}%
  \addtocounter{footnote}{-1}%
  \endgroup
}

\newcommand{\myparagraph}[1]{{\em #1}}

\newcommand{\newparentheses}[3]{%
  \expandafter\newcommand\csname #1\endcsname[1]{#2##1#3}%
  \expandafter\newcommand\csname #1L\endcsname[1]{\bigl#2##1\bigr#3}%
  \expandafter\newcommand\csname #1XL\endcsname[1]{\Bigl#2##1\Bigr#3}%
  \expandafter\newcommand\csname #1V\endcsname[1]{\left#2##1\right#3}}

\newparentheses{parens}{(}{)}
\newparentheses{braces}{\{}{\}}
\newparentheses{brackets}{[}{]}
\newparentheses{floor}{\lfloor}{\rfloor}
\newparentheses{ceil}{\lceil}{\rceil}
\newparentheses{abs}{|}{|}
\newparentheses{set}{\{}{\}}
\newparentheses{size}{|}{|}
\newparentheses{seq}{\langle}{\rangle}

\makeatletter
\newcommand{\onenewattribute}[4]{%
  \@ifundefined{#2}{\let\@@def\newcommand}{\let\@@def\renewcommand}%
  \expandafter\@@def\csname #2\endcsname[1][]{%
    \def\first@arg{##1}\csname @#2\endcsname}%
  \@ifundefined{@#2}{\let\@@def\newcommand}{\let\@@def\renewcommand}%
  \expandafter\@@def\csname @#2\endcsname[2][]{%
    \ifthenelse{\equal{#1}{sub}}%
    {\csname @@#2\endcsname{##1}{\first@arg}{##2}}%
    {\csname @@#2\endcsname{\first@arg}{##1}{##2}}}
  \@ifundefined{@@#2}{\let\@@def\newcommand}{\let\@@def\renewcommand}%
  \expandafter\@@def\csname @@#2\endcsname[3]{%
    \ifthenelse{\equal{##1}{}}%
    {\ifthenelse{\equal{##2}{}}%
      {#3\csname #4\endcsname{##3}}%
      {#3_{##2}\csname #4\endcsname{##3}}}%
    {\ifthenelse{\equal{##2}{}}%
      {#3^{##1}\csname #4\endcsname{##3}}%
      {#3_{##2}^{##1}\csname #4\endcsname{##3}}}}}
\newcommand{\newattribute}[3][sub]{%
  \onenewattribute{#1}{#2}{#3}{parens}%
  \onenewattribute{#1}{#2L}{#3}{parensL}%
  \onenewattribute{#1}{#2XL}{#3}{parensXL}%
  \onenewattribute{#1}{#2V}{#3}{parensV}}

\newcommand{\newproperty}[3][sub]{%
  \@ifundefined{#2}{\let\@@def\newcommand}{\let\@@def\renewcommand}%
  \expandafter\@@def\csname #2\endcsname[2][]{%
    \ifthenelse{\equal{#1}{sub}}%
    {\ifthenelse{\equal{##1}{}}%
      {#3_{##2}}%
      {#3_{##2}^{##1}}}%
    {\ifthenelse{\equal{##1}{}}%
      {#3^{##2}}%
      {#3_{##1}^{##2}}}}}
\makeatother

\newattribute{OhOf}{\mathcal{O}}
\newattribute{ThetaOf}{\Theta}
\newattribute{OmegaOf}{\Omega}
\newattribute{ohOf}{\mathcal{o}}
\newattribute{omegaOf}{\omega}

\newattribute{sort}{\mathrm{sort}}
\newattribute{psort}{\mathrm{sort}_P}
\newattribute{onesort}{\mathrm{sort}_1}
\newattribute{scan}{\mathrm{scan}}

\newattribute{scc}{\textrm{scc}}
\newattribute{prd}{\textrm{prd}}
\newcommand{\M}{\mathcal{M}}

\def\comment#1{}
\def\withcomments{%
  \addtolength{\oddsidemargin}{-0.5in}%
  \addtolength{\evensidemargin}{-0.5in}%
  \setlength{\marginparwidth}{1in}
  \newcounter{mycommentcounter}%
  \def\comment##1{\refstepcounter{mycommentcounter}%
    \ifhmode
      \unskip
      {\dimen1=\baselineskip
        \divide\dimen1 by 2%
        \raise\dimen1\llap{\tiny -\themycommentcounter-}}%
    \fi
    \marginpar{\renewcommand{\baselinestretch}{0.8}%
      \footnotesize [\themycommentcounter]: \raggedright ##1}}%
  \date{\framebox{Draft of \today}}}



%% file: abstract.tex
\begin{abstract}
In this paper we present a framework for designing algorithms in shared
memory of GPUs without incurring memory bank conflicts. 
Using our framework we develop the first comparison-based shared memory sorting
algorithm that incurs no bank conflicts. It can be used as a subroutine for GPU
sorting algorithms to replace current use of sorting networks in shared memory.
Using our bank conflict free
shared memory sorting subroutine as a black box, we design BCFMergesort, an algorithm for
merging sorted streams of data that are larger than shared memory.  Our
algorithm performs all accesses to global memory in coalesced manner and incurs
no bank conflicts during the merge. 
\blfootnote{Work 
supported in part by the DFG grant ME 2088/3-1 within the
priority programme SPP 1736 (Algorithms for Big Data) and by the
Danish National Research Foundation grant DNRF84 through Center for
Massive Data Algorithmics (MADALGO).}
\end{abstract}

%% file: intro.tex
\section{Introduction}
During the past decade GPUs -- the massively parallel processors developed for
graphics rendering -- have been adapted for general purpose computation and 
CUDA and OpenCL -- extensions for C programming language -- have been developed 
for programming them.  With hundreds of cores capable of running thousands of 
threads, GPUs have become a standard computational platform in high-performance 
computing (HPC) and have been successfully used to analyze data in natural 
sciences, such as biology, chemistry, physics and astronomy.  



\subsection{Brief introduction to GPU organization} 
\label{sec:intro:orga}
To manage the large number of cores, GPUs are designed hierarchically. They 
consist of the GPU's on-board memory, known as {\em global memory}, and a 
number of {\em streaming multiprocessors (SMs)}. Each SM consists of multiple 
{\em scalar processors}, or {\em cores}, and several types of faster but 
smaller memories.

Each physical core can support multiple threads. The management of so many 
threads comes with a very specific thread organization.  All threads on a GPU are 
organized into {\em warps} -- a collection of 32 threads -- which must 
execute code as a {\em single-instruction, multiple data (SIMD)} processor in 
an implicitly synchronized (lock-step) manner. SIMD execution implies that any 
if statement that requires some subset of threads to execute one instruction, 
while another subset to execute another instruction, will result in the two 
subsets of threads waiting for each other. This is known as {\em branch 
divergence} in GPU literature and can result in drastic loss of parallelism for 
nested conditional statements. 

Warps are grouped into {\em cooperative thread arrays (CTAs)}, also known as 
{\em thread blocks}. All warps of a CTA must execute the same code, are run 
asynchronously from each other with the ability to call explicit barrier 
synchronizations. All warps of a CTA are guaranteed to be run on the same 
physical SM. Finally, synchronization across multiple CTAs can be performed 
using explicit barriers.  

The memory of a GPU is organized into multiple memory types each with its own 
advantages, but also their own challenges. 

\begin{itemize}
\item {\em Global memory} is the largest but the slowest of all memory types.
To hide the latency associated with access to global memory, GPU programmers
are encouraged to {\em coalesce} their accesses to global memory.  The accesses
are coalesced when the threads of a warp access contiguous address space in the
global memory. Global memory is accessible by all threads running on the GPU at
all times and the programmer is responsible to ensure the correctness of the
data under concurrent updates in such an asynchronous setting. 
\item {\em Shared memory} is faster but also smaller than global memory. It 
does not require coalesced accesses, however, it is organized into a limited 
number of memory banks. Each thread can access any memory bank, however, 
simultaneous accesses to the same memory bank by multiple threads of a warp 
cause a contention, known in GPU literature as {\em bank conflicts}. The 
threads causing a bank conflict are serviced sequentially resulting in the loss 
of parallelism. Each CTA implements its own address space within shared memory 
and the data in shared memory is exclusive to all threads of a single CTA, 
i.e., the threads of one CTA cannot access the data of another CTA via shared 
memory. 
\item {\em Registers} are the fastest type of memory on GPUs. Efficient access 
to data stored in registers does not require coalescing, nor does it suffer 
from memory bank conflicts. However, registers are private to each thread, very 
limited in size (only 63 of them per thread on our NVIDIA GTX 580 graphics 
card) and the addressing must be exactly the same for all threads of a warp. If 
more registers are required by a thread, the excess is stored in {\em local 
memory}. 
\item {\em Local memory} is not a separate physical memory type, but instead is
implemented as a logical partition of global memory. However, the latency
of accessing slow global memory is mitigated by automatic caching of recently
used data of local memory in the same memory banks where shared memory resides. 
%
\item {\em Texture memory} and {\em Constant memory} are other types of memory 
which are very useful in graphics rendering applications. However they are very 
specialized and their use for general purpose computations is very poorly 
understood.  In this paper we do not use those memories. 
\end{itemize}

\subsection{Effects of bank conflicts on runtime}

Due to large latency of accessing global memory, fast practical implementations 
typically first load data into shared memory and process data from there. Efficient 
access to contiguous data in global memory through coalescing is well understood 
and Merrill and Grimshaw~\cite{merrill:scan} has optimized the process to get 
sustainable throughput that are close to theoretical maximal.

However, once the data in shared memory, the access patterns in the presence of 
banked memory is less understood. 
%
%
%

To observe the effects of bank conflicts on runtime, consider the problem of {\em colored prefix sums}:  given an array of $N$ elements 
and a set of $d$ colors $\{c_1, c_2, \dots, c_d\}$, with a color $c_i$ associated 
with each element, colored prefix sums asks to compute $d$ independent prefix sums 
among the elements of the same color.

Given $P \ll N$ processors, colored prefix sums can be solved using the standard three phase approach to solving prefix sums: 
partition the input into $P$ subsets of contiguous $N/P$ elements and 
compute the $d$ sums (one for each color) within each subset using an independent processor; compute $d$ independent prefix 
sums on the $d$ sets of $P$ values each from previous phase; finally add the value of the appropriate 
color from the previous step while computing the prefix sum within the subsets for 
each of $d$ colors using independent processor per subset.
%
If $P \ll N$ the runtime of the second step is negligible (in 
practice it is typically performed using a single processor) and the algorithm is 
dominated by the first and the third step of the algorithm. If the second step 
takes negligible time and $d \ll N/P$, the runtime should vary little as a function 
of $d$. 

\begin{figure*}[tb]
\subfloat[]{
  \includegraphics[width=1.65in, angle=270]{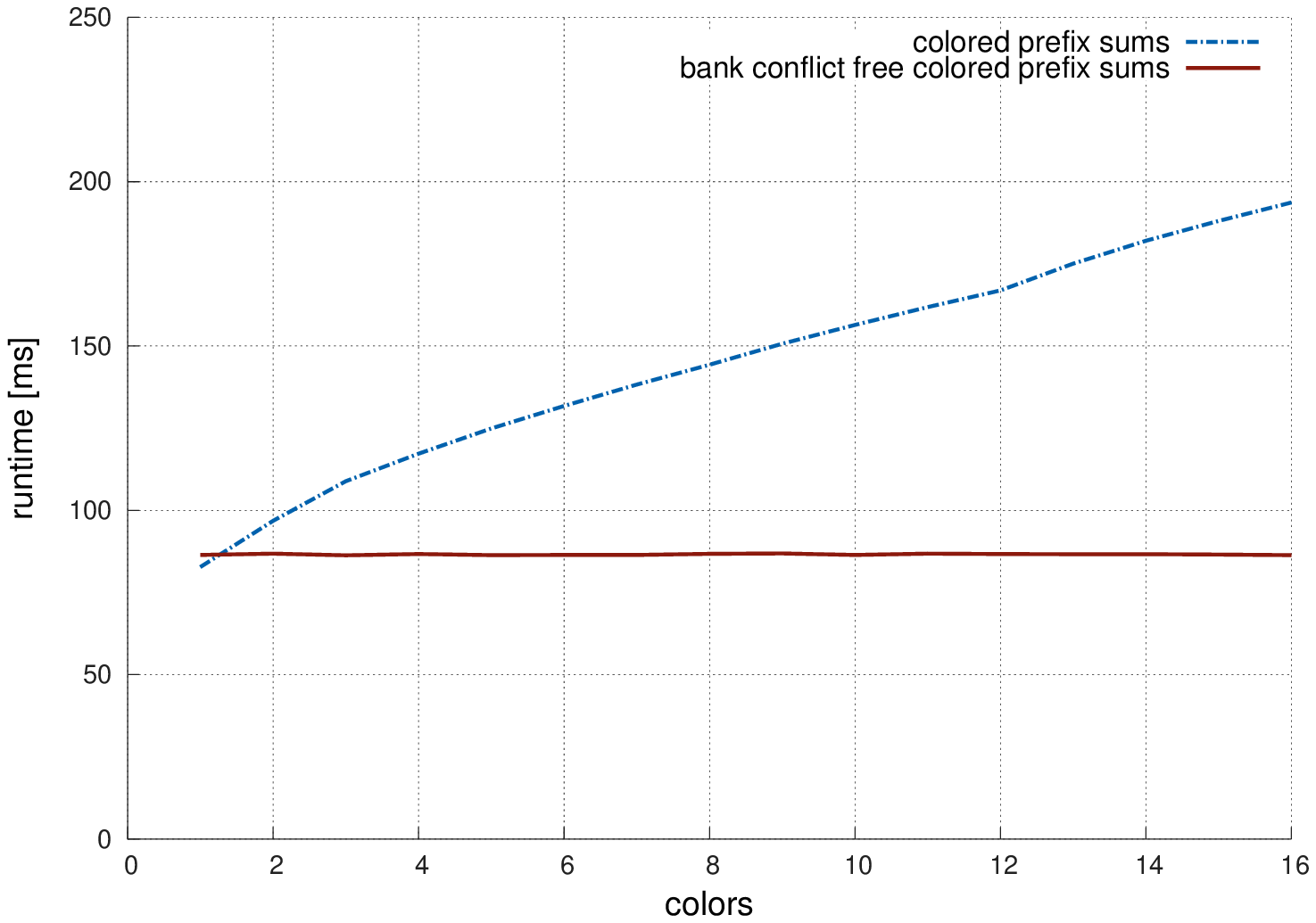}
  \label{fig:cps-runtime-a}
}
~
\subfloat[]{
  \includegraphics[width=1.65in, angle=270]{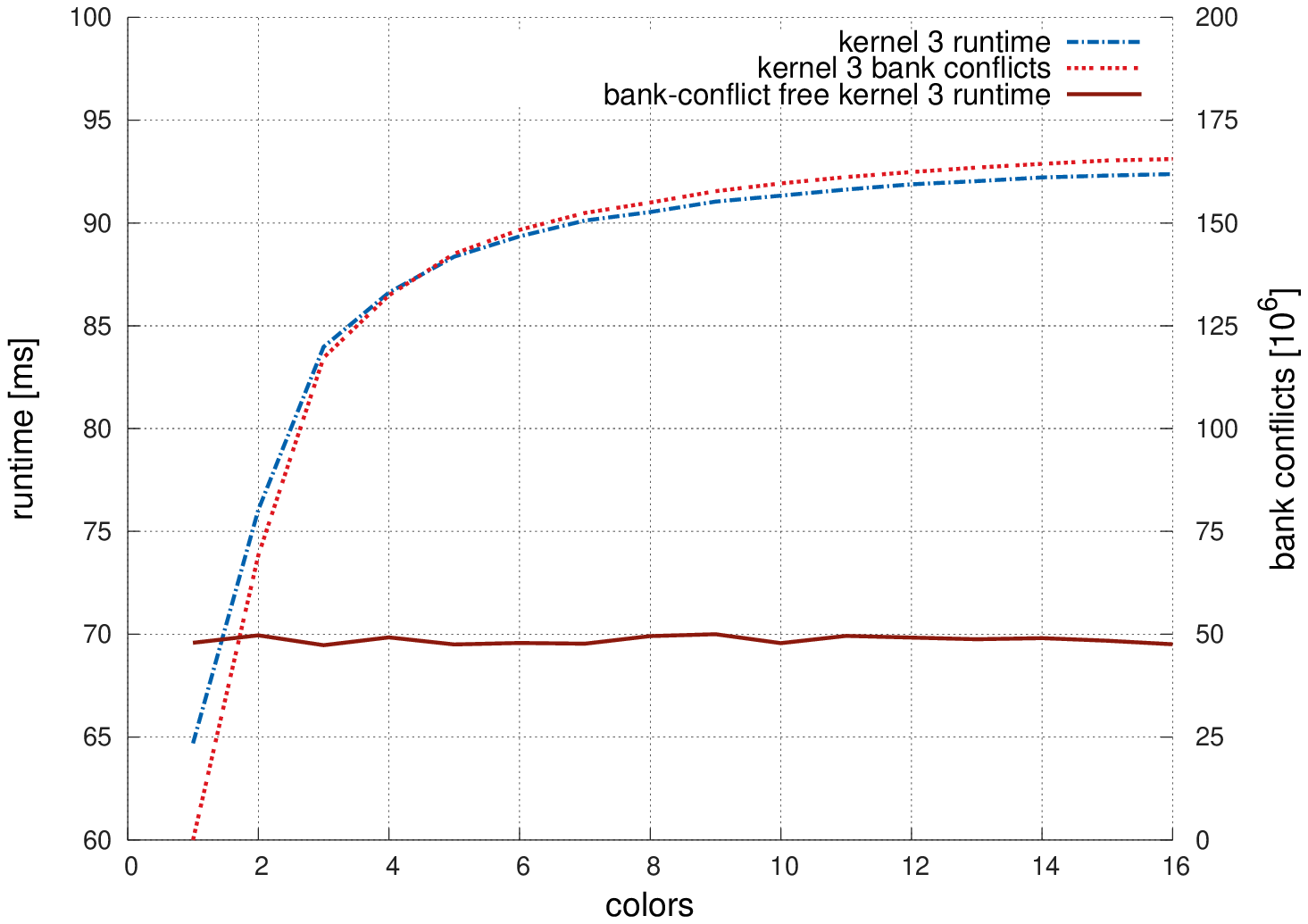}
  \label{fig:cps-runtime-b}
}
\caption{The runtime of the colored prefix sums algorithm rises with increasing 
		number of colors (a). The correlation between the number of bank conflicts 
		and runtime can be seen in the example of the prefix sums kernel (b). The 
		input size is $2^{29}$ elements (2~GB).}
\label{fig:cps-runtime}
\end{figure*}

Figure~\ref{fig:cps-runtime-a} (dashed line) shows the actual runtime of colored prefix sums. 
Unlike our prediction, we can see that it grows significantly as a function of 
$d$. Figure~\ref{fig:cps-runtime-b} shows the runtime of just the third phase 
and the number of bank conflicts incurred during execution. One can clearly see 
a correlation between the two. After implemented an improved version of colored 
prefix sums that does not incur any bank conflicts (see 
Appendix~\ref{sec:colored-prefix-sums} for implementation details), one can see that the runtime 
of the third phase indeed becomes constant as a function of $d$ (solid lines). This simple example demonstrates that if we want to be able 
to accurately predict runtimes of GPU algorithms using asymptotic  analysis, it 
is very important to design algorithms without any bank conflicts.



\subsection{Past work on sorting on GPUs} 
Sorting numbers in an array was one of the first combinatorial problems 
implemented on GPUs. Purcel et al.~\cite{purcell2003photon} were the first to 
implement a bitonic sorting network on a GPU, while Sintorn et 
al.~\cite{sintorn2008fast} implemented the first sorting algorithm in the CUDA 
programming environment. Since then many different algorithms have been 
implemented on GPUs, e.g. 
mergesort~\cite{satish:mergesort,davidson2012efficient,ye:sort}, 
quicksort~\cite{cederman2008practical}, bitonic sort~\cite{peters2010fast}, 
radixsort~\cite{merrill2011high} and samplesort~\cite{gpu-sample-sort}.

For comparison-based algorithms, when data is small enough, fast
implementations typically implement a sorting network (usually Batcher's 
odd-even mergesort or bitonic mergesort). The likely reason for the success of 
sorting networks on GPUs is the small constant factors in the run time analysis 
of sorting networks and their data-oblivious nature. The data-obliviousness of 
sorting networks implies that the execution stream of every thread is the same 
regardless of the value of the input data. Thus, there is no need for 
conditional statements in the code which bodes well with the SIMD nature of 
threads in a warp.  

However, the rigid access pattern of the sorting networks makes it difficult to 
avoid memory bank conflicts and over the decade of GPU computing no one has 
been able to come up with an efficient sorting network that causes no bank 
conflicts on inputs larger than the number of memory banks in shared memory.

\subsection{Our Contributions}
We present a framework for designing bank conflict free algorithms in shared 
memory. Using our framework we identified an existing 27 year old sorting algorithm as a 
candidate for a bank conflict free comparison-based sorting algorithm on GPUs for data 
as large as shared memory. 
%
%
%
%

Using our new sorting algorithm as a subroutine, we developed a bank conflict 
free merging of pairs of sorted sequences, thus, producing {\em BCFMergesort} 
-- the first fully bank conflicts free comparison-based sorting GPU algorithm. 

The lack of bank conflicts in BCFMergesort allows us to perform
a worst-case theoretical analysis and accurately predict the runtime of our
algorithm on GPUs. This is the first comparison-based sorting algorithm on GPUs
that we are aware of that has such theoretical analysis. 
We implemented our algorithm and show that it outperforms current GPU mergesort 
implementations.

%% file: framework.tex
\section{Algorithmic Framework}
\label{sec:framework}
Let $P_h$ be the number of SMs on the hardware and $M_h$ be the amount of shared 
memory available on each SM. Let $P$ be the total number of CTAs. Let $w$ be the 
number of threads within each warp and for simplicity we let $w$ represent the 
number of banks in shared memory. Let $M$ be the size of shared memory required 
by each CTA. We choose $M$ to be at most $\frac{P_h\cdot M_h}{P}$ to guarantee that all CTAs 
can fit on all SMs and run concurrently without any context switches.  Let $W$ 
be the number of warps per CTA.  We set $P$ and $W$ to be the minimum values 
which maximize the throughput between global and shared memories. For ease of 
exposition we define $p = PW$. In our experiments we observed that 
$p = PW = 8P_h$ was the optimal choice for $P$ and $W$ to maximize throughput. 
This relationship and the constraint $M \le \frac{P_h\cdot M_h}{P}$ allows us to vary $M$ 
depending on the specific needs of the algorithm. 

During algorithm design, we treat $\ThetaOf{M}$ elements as a single contiguous 
block and transfer these blocks between the global and shared memories. Note 
that $M > 32$, is the lower bound to perform such transfer in coalesced manner. 
Our experiments showed that loading $M \ge 1024$ elements per warp maximizes the 
throughput. 
Once $M$ elements are in shared memory, we can focus on processing them without 
any bank conflicts as efficiently as possible.

\input{bank_conflicts2}

%% file: bank_conflicts2.tex
\subsection{Designing Bank Conflict Free Algorithms In Shared Memory}
\label{sec:bank-conflicts}

\begin{figure*}[tb]
\begin{center}
\begin{tikzpicture}[scale=.18,auto,every node/.style={inner sep=0pt,outer sep=0pt,circle,minimum size=10pt,font=\tiny}]
\def\horizstart{5}
\def\horizspace{3}
\def\vertstart{0}
\def\vertspace{2}
\def\labelstart{-1}
\useasboundingbox (2*\labelstart-\horizstart,\vertstart) rectangle (\horizstart+13*\horizspace,\vertstart+8*\vertspace);
\definecolor{memcolor}{RGB}{255,189,0}
\fill[color=memcolor] (\horizstart,\vertstart) rectangle (\horizstart+13*\horizspace,\vertstart+8*\vertspace);
\coordinate  (aa)  at  (\horizstart,\vertstart+0*\vertspace);
\coordinate  (ab)  at  (\horizstart+13*\horizspace,\vertstart+0*\vertspace);
\coordinate  (ba)  at  (\horizstart,\vertstart+1*\vertspace);
\coordinate  (bb)  at  (\horizstart+13*\horizspace,\vertstart+1*\vertspace);
\coordinate  (ca)  at  (\horizstart,\vertstart+2*\vertspace);
\coordinate  (cb)  at  (\horizstart+13*\horizspace,\vertstart+2*\vertspace);
\coordinate  (da)  at  (\horizstart,\vertstart+3*\vertspace);
\coordinate  (db)  at  (\horizstart+13*\horizspace,\vertstart+3*\vertspace);
\coordinate  (ea)  at  (\horizstart,\vertstart+4*\vertspace);
\coordinate  (eb)  at  (\horizstart+13*\horizspace,\vertstart+4*\vertspace);
\coordinate  (fa)  at  (\horizstart,\vertstart+5*\vertspace);
\coordinate  (fb)  at  (\horizstart+13*\horizspace,\vertstart+5*\vertspace);
\coordinate  (ga)  at  (\horizstart,\vertstart+6*\vertspace);
\coordinate  (gb)  at  (\horizstart+13*\horizspace,\vertstart+6*\vertspace);
\coordinate  (ha)  at  (\horizstart,\vertstart+7*\vertspace);
\coordinate  (hb)  at  (\horizstart+13*\horizspace,\vertstart+7*\vertspace);
\coordinate  (ia)  at  (\horizstart,\vertstart+8*\vertspace);
\coordinate  (ib)  at  (\horizstart+13*\horizspace,\vertstart+8*\vertspace);
\coordinate  (ma)  at  (\horizstart+1*\horizspace,\vertstart);
\coordinate  (mb)  at  (\horizstart+1*\horizspace,\vertstart+8*\vertspace);
\coordinate  (na)  at  (\horizstart+2*\horizspace,\vertstart);
\coordinate  (nb)  at  (\horizstart+2*\horizspace,\vertstart+8*\vertspace);
\coordinate  (oa)  at  (\horizstart+3*\horizspace,\vertstart);
\coordinate  (ob)  at  (\horizstart+3*\horizspace,\vertstart+8*\vertspace);
\coordinate  (pa)  at  (\horizstart+4*\horizspace,\vertstart);
\coordinate  (pb)  at  (\horizstart+4*\horizspace,\vertstart+8*\vertspace);
\coordinate  (qa)  at  (\horizstart+5*\horizspace,\vertstart);
\coordinate  (qb)  at  (\horizstart+5*\horizspace,\vertstart+8*\vertspace);
\coordinate  (ra)  at  (\horizstart+6*\horizspace,\vertstart);
\coordinate  (rb)  at  (\horizstart+6*\horizspace,\vertstart+8*\vertspace);
\coordinate  (sa)  at  (\horizstart+7*\horizspace,\vertstart);
\coordinate  (sb)  at  (\horizstart+7*\horizspace,\vertstart+8*\vertspace);
\coordinate  (ta)  at  (\horizstart+8*\horizspace,\vertstart);
\coordinate  (tb)  at  (\horizstart+8*\horizspace,\vertstart+8*\vertspace);
\coordinate  (ua)  at  (\horizstart+9*\horizspace,\vertstart);
\coordinate  (ub)  at  (\horizstart+9*\horizspace,\vertstart+8*\vertspace);
\coordinate  (va)  at  (\horizstart+10*\horizspace,\vertstart);
\coordinate  (vb)  at  (\horizstart+10*\horizspace,\vertstart+8*\vertspace);
\coordinate  (wa)  at  (\horizstart+11*\horizspace,\vertstart);
\coordinate  (wb)  at  (\horizstart+11*\horizspace,\vertstart+8*\vertspace);
\coordinate  (xa)  at  (\horizstart+12*\horizspace,\vertstart);
\coordinate  (xb)  at  (\horizstart+12*\horizspace,\vertstart+8*\vertspace);
\draw[black]  (aa)  --  (ab);
\draw[black]  (ba)  --  (bb);
\draw[black]  (ca)  --  (cb);
\draw[black]  (da)  --  (db);
\draw[black]  (ea)  --  (eb);
\draw[black]  (fa)  --  (fb);
\draw[black]  (ga)  --  (gb);
\draw[black]  (ha)  --  (hb);
\draw[black]  (ia)  --  (ib);
\draw[black]  (aa)  --  (ia);
\draw[black]  (ma)  --  (mb);
\draw[black]  (na)  --  (nb);
\draw[black]  (oa)  --  (ob);
\draw[black]  (pa)  --  (pb);
\draw[black]  (qa)  --  (qb);
\draw[black]  (ra)  --  (rb);
\draw[black]  (sa)  --  (sb);
\draw[black]  (ta)  --  (tb);
\draw[black]  (ua)  --  (ub);
\draw[black]  (va)  --  (vb);
\draw[black]  (wa)  --  (wb);
\draw[black]  (xa)  --  (xb);
\draw[black]  (ab)  --  (ib);
\node[draw=none,fill=none] (mb0) at (\labelstart,\vertstart+0.5*\vertspace)   {Memory Bank 7};
\node[draw=none,fill=none] (mb1) at (\labelstart,\vertstart+1.5*\vertspace)   {Memory Bank 6};
\node[draw=none,fill=none] (mb2) at (\labelstart,\vertstart+2.5*\vertspace)   {Memory Bank 5};
\node[draw=none,fill=none] (mb3) at (\labelstart,\vertstart+3.5*\vertspace)   {Memory Bank 4};
\node[draw=none,fill=none] (mb4) at (\labelstart,\vertstart+4.5*\vertspace)   {Memory Bank 3};
\node[draw=none,fill=none] (mb5) at (\labelstart,\vertstart+5.5*\vertspace)   {Memory Bank 2};
\node[draw=none,fill=none] (mb6) at (\labelstart,\vertstart+6.5*\vertspace)   {Memory Bank 1};
\node[draw=none,fill=none] (mb7) at (\labelstart,\vertstart+7.5*\vertspace)   {Memory Bank 0};
\node[draw=none,fill=none] (a0) at (\horizstart+0.5*\horizspace,\vertstart+0.5*\vertspace)   {$A_{7}$};
\node[draw=none,fill=none] (a1) at (\horizstart+0.5*\horizspace,\vertstart+1.5*\vertspace)   {$A_{6}$};
\node[draw=none,fill=none] (a2) at (\horizstart+0.5*\horizspace,\vertstart+2.5*\vertspace)   {$A_{5}$};
\node[draw=none,fill=none] (a3) at (\horizstart+0.5*\horizspace,\vertstart+3.5*\vertspace)   {$A_{4}$};
\node[draw=none,fill=none] (a4) at (\horizstart+0.5*\horizspace,\vertstart+4.5*\vertspace)   {$A_{3}$};
\node[draw=none,fill=none] (a5) at (\horizstart+0.5*\horizspace,\vertstart+5.5*\vertspace)   {$A_{2}$};
\node[draw=none,fill=none] (a6) at (\horizstart+0.5*\horizspace,\vertstart+6.5*\vertspace)   {$A_{1}$};
\node[draw=none,fill=none] (a7) at (\horizstart+0.5*\horizspace,\vertstart+7.5*\vertspace)   {$A_{0}$};
\node[draw=none,fill=none] (a8) at (\horizstart+1.5*\horizspace,\vertstart+0.5*\vertspace)   {$A_{15}$};
\node[draw=none,fill=none] (a9) at (\horizstart+1.5*\horizspace,\vertstart+1.5*\vertspace)   {$A_{14}$};
\node[draw=none,fill=none] (a10) at (\horizstart+1.5*\horizspace,\vertstart+2.5*\vertspace)   {$A_{13}$};
\node[draw=none,fill=none] (a11) at (\horizstart+1.5*\horizspace,\vertstart+3.5*\vertspace)   {$A_{12}$};
\node[draw=none,fill=none] (a12) at (\horizstart+1.5*\horizspace,\vertstart+4.5*\vertspace)   {$A_{11}$};
\node[draw=none,fill=none] (a13) at (\horizstart+1.5*\horizspace,\vertstart+5.5*\vertspace)   {$A_{10}$};
\node[draw=none,fill=none] (a14) at (\horizstart+1.5*\horizspace,\vertstart+6.5*\vertspace)   {$A_{9}$};
\node[draw=none,fill=none] (a15) at (\horizstart+1.5*\horizspace,\vertstart+7.5*\vertspace)   {$A_{8}$};
\node[draw=none,fill=none] (a16) at (\horizstart+2.5*\horizspace,\vertstart+0.5*\vertspace)   {$A_{23}$};
\node[draw=none,fill=none] (a17) at (\horizstart+2.5*\horizspace,\vertstart+1.5*\vertspace)   {$A_{22}$};
\node[draw=none,fill=none] (a18) at (\horizstart+2.5*\horizspace,\vertstart+2.5*\vertspace)   {$A_{21}$};
\node[draw=none,fill=none] (a19) at (\horizstart+2.5*\horizspace,\vertstart+3.5*\vertspace)   {$A_{20}$};
\node[draw=none,fill=none] (a20) at (\horizstart+2.5*\horizspace,\vertstart+4.5*\vertspace)   {$A_{19}$};
\node[draw=none,fill=none] (a21) at (\horizstart+2.5*\horizspace,\vertstart+5.5*\vertspace)   {$A_{18}$};
\node[draw=none,fill=none] (a22) at (\horizstart+2.5*\horizspace,\vertstart+6.5*\vertspace)   {$A_{17}$};
\node[draw=none,fill=none] (a23) at (\horizstart+2.5*\horizspace,\vertstart+7.5*\vertspace)   {$A_{16}$};
\node[draw=none,fill=none] (a24) at (\horizstart+3.5*\horizspace,\vertstart+0.5*\vertspace)   {$A_{31}$};
\node[draw=none,fill=none] (a25) at (\horizstart+3.5*\horizspace,\vertstart+1.5*\vertspace)   {$A_{30}$};
\node[draw=none,fill=none] (a26) at (\horizstart+3.5*\horizspace,\vertstart+2.5*\vertspace)   {$A_{29}$};
\node[draw=none,fill=none] (a27) at (\horizstart+3.5*\horizspace,\vertstart+3.5*\vertspace)   {$A_{28}$};
\node[draw=none,fill=none] (a28) at (\horizstart+3.5*\horizspace,\vertstart+4.5*\vertspace)   {$A_{27}$};
\node[draw=none,fill=none] (a29) at (\horizstart+3.5*\horizspace,\vertstart+5.5*\vertspace)   {$A_{26}$};
\node[draw=none,fill=none] (a30) at (\horizstart+3.5*\horizspace,\vertstart+6.5*\vertspace)   {$A_{25}$};
\node[draw=none,fill=none] (a31) at (\horizstart+3.5*\horizspace,\vertstart+7.5*\vertspace)   {$A_{24}$};
\node[draw=none,fill=none] (a32) at (\horizstart+4.5*\horizspace,\vertstart+0.5*\vertspace)   {$A_{39}$};
\node[draw=none,fill=none] (a33) at (\horizstart+4.5*\horizspace,\vertstart+1.5*\vertspace)   {$A_{38}$};
\node[draw=none,fill=none] (a34) at (\horizstart+4.5*\horizspace,\vertstart+2.5*\vertspace)   {$A_{37}$};
\node[draw=none,fill=none] (a35) at (\horizstart+4.5*\horizspace,\vertstart+3.5*\vertspace)   {$A_{36}$};
\node[draw=none,fill=none] (a36) at (\horizstart+4.5*\horizspace,\vertstart+4.5*\vertspace)   {$A_{35}$};
\node[draw=none,fill=none] (a37) at (\horizstart+4.5*\horizspace,\vertstart+5.5*\vertspace)   {$A_{34}$};
\node[draw=none,fill=none] (a38) at (\horizstart+4.5*\horizspace,\vertstart+6.5*\vertspace)   {$A_{33}$};
\node[draw=none,fill=none] (a39) at (\horizstart+4.5*\horizspace,\vertstart+7.5*\vertspace)   {$A_{32}$};
\node[draw=none,fill=none] (a40) at (\horizstart+5.5*\horizspace,\vertstart+0.5*\vertspace)   {$A_{47}$};
\node[draw=none,fill=none] (a41) at (\horizstart+5.5*\horizspace,\vertstart+1.5*\vertspace)   {$A_{46}$};
\node[draw=none,fill=none] (a42) at (\horizstart+5.5*\horizspace,\vertstart+2.5*\vertspace)   {$A_{45}$};
\node[draw=none,fill=none] (a43) at (\horizstart+5.5*\horizspace,\vertstart+3.5*\vertspace)   {$A_{44}$};
\node[draw=none,fill=none] (a44) at (\horizstart+5.5*\horizspace,\vertstart+4.5*\vertspace)   {$A_{43}$};
\node[draw=none,fill=none] (a45) at (\horizstart+5.5*\horizspace,\vertstart+5.5*\vertspace)   {$A_{42}$};
\node[draw=none,fill=none] (a46) at (\horizstart+5.5*\horizspace,\vertstart+6.5*\vertspace)   {$A_{41}$};
\node[draw=none,fill=none] (a47) at (\horizstart+5.5*\horizspace,\vertstart+7.5*\vertspace)   {$A_{40}$};
\node[draw=none,fill=none] (a48) at (\horizstart+6.5*\horizspace,\vertstart+0.5*\vertspace)   {$A_{55}$};
\node[draw=none,fill=none] (a49) at (\horizstart+6.5*\horizspace,\vertstart+1.5*\vertspace)   {$A_{54}$};
\node[draw=none,fill=none] (a50) at (\horizstart+6.5*\horizspace,\vertstart+2.5*\vertspace)   {$A_{53}$};
\node[draw=none,fill=none] (a51) at (\horizstart+6.5*\horizspace,\vertstart+3.5*\vertspace)   {$A_{52}$};
\node[draw=none,fill=none] (a52) at (\horizstart+6.5*\horizspace,\vertstart+4.5*\vertspace)   {$A_{51}$};
\node[draw=none,fill=none] (a53) at (\horizstart+6.5*\horizspace,\vertstart+5.5*\vertspace)   {$A_{50}$};
\node[draw=none,fill=none] (a54) at (\horizstart+6.5*\horizspace,\vertstart+6.5*\vertspace)   {$A_{49}$};
\node[draw=none,fill=none] (a55) at (\horizstart+6.5*\horizspace,\vertstart+7.5*\vertspace)   {$A_{48}$};
\node[draw=none,fill=none] (a56) at (\horizstart+7.5*\horizspace,\vertstart+0.5*\vertspace)   {$A_{63}$};
\node[draw=none,fill=none] (a57) at (\horizstart+7.5*\horizspace,\vertstart+1.5*\vertspace)   {$A_{62}$};
\node[draw=none,fill=none] (a58) at (\horizstart+7.5*\horizspace,\vertstart+2.5*\vertspace)   {$A_{61}$};
\node[draw=none,fill=none] (a59) at (\horizstart+7.5*\horizspace,\vertstart+3.5*\vertspace)   {$A_{60}$};
\node[draw=none,fill=none] (a60) at (\horizstart+7.5*\horizspace,\vertstart+4.5*\vertspace)   {$A_{59}$};
\node[draw=none,fill=none] (a61) at (\horizstart+7.5*\horizspace,\vertstart+5.5*\vertspace)   {$A_{58}$};
\node[draw=none,fill=none] (a62) at (\horizstart+7.5*\horizspace,\vertstart+6.5*\vertspace)   {$A_{57}$};
\node[draw=none,fill=none] (a63) at (\horizstart+7.5*\horizspace,\vertstart+7.5*\vertspace)   {$A_{56}$};
\end{tikzpicture}
\caption{The view of shared memory as a $b \times \ceil{M/b}$ matrix.}
\label{fig:shared-memory}
\end{center}
\end{figure*}
%

We extend the view of shared memory as observed by Dotsenko et
al.~\cite{dotsenko:scan}. Just like them we view the shared memory of size $M$
as a matrix $\M$, with $w$ rows and $\ceilV{\frac{M}{w}}$ columns. Each row corresponds to one
of the $w$ memory banks. When $r$ items are loaded into contiguous address space
in shared memory, it is loaded in column major order into $\ceilV{\frac{r}{w}}$ columns (see
Figure~\ref{fig:shared-memory}).  Dotsenko et al. used this view to pad the
input to offset the starting indices of each thread, so that a parallel scan of
contiguous data would result in each thread starting from a different memory
bank. 

We extend this view to computations other than parallel scan. In particular,
observe that if we view the shared memory as a matrix and process it in such a
way that each thread processes a separate row of the matrix, each thread will be
accessing a separate bank, thus resulting in no bank conflicts. Furthermore,
observe that if we can transform the matrix from the column-major layout to the 
row-major layout without causing any bank conflicts, the original columns become
rows and, therefore, the columns of the original matrix can also be processed without
any bank conflicts. 
A transformation of a square
matrix between the two layouts is simply a transposition of it. A warp can perform bank conflict
free transposition of a $w \times w$ matrix in $\frac{w}{2}-1$ rounds,
where in each round $i$: $1 \le i \le \frac{w}{2}-1$, each thread $0 \le t \le w-1$ swaps the
element $\M[t, (i+t)\mod w]$ with the element $\M[(i+t)\mod w, t]$.  Note that
within each reading or writing step, all $w$ threads access a different row
(memory bank) of the matrix. The lock-step execution among the threads of
the warp ensures that the reads and writes are performed one after another
without interfering with each other. 

Since the transposition incurs no bank conflicts, all $w$ threads can perform each
step without interfering with each other and we can easily analyze the
asymptotic worst-case runtime as $\left(\frac{w}{2}-1\right)\cdot\OhOf{1} = \OhOf{w}$ parallel steps.

We can easily generalize the transformation for input sizes $n = hw^2$ for a small integer constant $h$ that divides $w$. The details are presented in Appendix~\ref{sec:long-shearsort}.

%% file: sorting.tex
\section{BFCMergesort: Bank-conflict free sorting}
\label{sec:sorting}

For our bank conflict free sorting we chose the classic mergesort algorithm. Practical 
mergesort implementations distinguish two phases: the base case phase and the merging phase. 
In the base case phase, $\ceilV{\frac{N}{M}}$ sequences of $M$ elements each are sorted. The merging 
phase proceeds in $\ceilV{\log{\left(\frac{N}{M}\right)}}$ rounds, in each round merging pairs of sorted 
sequences from the previous round to create increasingly larger sequences, until only a 
single sorted sequence remains.  In this section, we present how to implement each of 
these phases without any bank conflicts.

\subsection{Base case (ShearSort)}
\label{sec:basecase}


On GPUs, the base case is sorted in shared memory using a sorting network. Sorting networks 
are ideal for GPUs because they are data oblivious (resulting in no branch divergence) 
and they are very simple (resulting in very small constant factors). However, 
sorting networks still cause bank conflicts if the size of the sequence to be sorted is larger than $w$. This 
is probably the reason why the fastest mergesort implementations on GPUs (e.g. the Thrust 
library~\cite{satish:mergesort}, the CUDPP library~\cite{davidson2012efficient} and 
warpsort~\cite{ye:sort}) use sorting networks only for up to $2w = 64$ elements.


In 1983, Sen et al.\cite{scherson:shear-sort} introduced ShearSort -- a sorting 
network to sort numbers arranged in an $m \times m$ matrix in snake-like order along the 
column-major order. The algorithm proceeds in rounds. In each round
it sorts even-numbered columns in ascending order and odd-numbered columns in descending 
order, then sorts each row in ascending order. Using 0-1 principle~\cite{knuth:sorting}, it 
is easy to show that after $\ThetaOf{\log m}$ rounds the matrix is sorted.

Using our observation from Section~\ref{sec:bank-conflicts}, it is easy to see that 
ShearSort can be implemented on GPUs to sort sequences of $M = w^2$ elements in shared memory 
with $w$ banks without any bank conflicts. Using $w$ threads, it will take 
$T_b(M) = \OhOfV{\hat{t}\left(\frac{M}{w}\right)\cdot \log w}$ parallel time, where $\hat{t}(n)$ is the time it takes to 
sort a row or a column of $n$ elements using a single thread.
We can still benefit from the simplicity and data-oblivious nature of traditional sorting networks by 
implementing sequential sorting of rows and columns with a sorting network. Thus, if the 
row/column sorting is implemented using Batcher's sorting network~\cite{batcher:sorting}, 
then $\hat{t}(n) = \OhOf{n\log^2 n}$ and the total time it takes to sort $M = w^2$ elements 
is $T_b(M) = \OhOf{\frac{M}{w}\log^3 w} = \OhOf{w\log^3 w} $. Notice, that although this approach incurs $\OhOf{\log w}$ extra 
steps over a simple implementation of a sorting network in shared memory, this time 
complexity is guaranteed in the worst case and our experiments show that the savings due to 
lack of bank conflicts are larger than the extra $\OhOf{\log w}$ factor (see 
Section~\ref{sec:experiments}).



The following simple extension allows us to sort larger inputs: for any integer $h$ that 
divides $w$, we can sort $M = w^2h$ elements using $h$ warps of a CTA. Remember that $w^2h$ elements can 
be viewed in shared memory as a $w \times (wh)$ matrix.  Each column of this matrix can be 
sorted as before. However, each row now consists of $wh$ elements and sorting it using a single thread would be too slow. Instead, 
we first pack $\frac{w}{h}$ such rows into separate $w \times w$ matrices as follows. 
We transpose each of $h$ square submatrices and gather the columns that comprise an 
original row of $wh$ elements into adjacent columns. These columns can easily be identified 
using index arithmetic. Once each original row is packed into $h$ contiguous columns, we 
run {\em Segmented ShearSort} on $w \times w$ submatrices. Segmented ShearSort is a modified 
ShearSort, which instead of sorting the whole rows of a square matrix, sorts groups of 
$\frac{w}{h}$ elements within each row. Thus, after $\log \left(\frac{w}{h}\right)$ rounds, 
each $w \times \left(\frac{w}{h}\right)$ submatrix corresponding to the original rows of 
$wh$ elements will be sorted.


Transformation of the $w\times (wh)$ matrices takes $\OhOf{w}$ parallel time when 
performed using $h$ warps ($wh$ threads). Segmented ShearSort of a square matrix 
using $w$ threads takes $\OhOfV{\hat{t}\left(\frac{M}{wh}\right) \log \frac{w}{h}} = \OhOf{ w\cdot \log^2 w \cdot \log \frac{w}{h}}$ parallel time, which is the total runtime for sorting $M = w^2h$ elements using $h$ warps.

%% file: merging.tex
\subsection{Bank-conflict free merging}
\label{sec:merging}

After sorting the base case, our algorithm performs a binary merge of the resulting sorted 
sequences using a separate warp to merge each pair of sequences. 
%
%

Let $P$ be the minimal number of CTAs that maximizes throughput, as mentioned in Section~\ref{sec:framework}. If the number of sequences to be merged is fewer than $2P$, some of the CTAs will 
be sitting idle, thus, reducing the available parallelism. Ye et al.~\cite{ye:sort} 
presented a randomized approach to mitigate this problem. Their idea is to suspend binary 
merging when the number of sequences drops below $2P$, distribute the sequences into $P$ 
buckets, such that all elements of bucket $i$ are smaller than elements in bucket $i+1$ for 
all buckets, and continue binary merging on the portions of the sequences within each 
bucket. Ye et al. choose bucket boundaries as random samples of the input.  In this section 
we present a deterministic approach for determining bucket boundaries, while adapting their 
merging algorithm to use our ShearSort from the previous section. 


\myparagraph{Binary merging.}
To merge two sorted sequences $A$ and $B$, a standard sequential algorithm starts 
off by loading the smallest element from each sequence. Then it repeatedly outputs 
the smaller of the two items and loads the next element from the sequence, whose 
element was written out.


Since each warp consists of $w=32$ threads, it is inefficient to load only one element from 
each sequence. Instead Ye et al~\cite{ye:sort} load a page of size $\frac{M}{2}$ from each 
sequence into shared memory, merge $M$ elements using all $w$ threads of a warp, output at 
least $\frac{M}{2}$ smallest elements, and load the next page of $\frac{M}{2}$ elements from 
the sequence, whose input is exhausted first.


%

In our implementation, we set $M = w^2$ and use our ShearSort implementation to merge the 
two pages in shared memory. Since both input sequences are already sorted, by the 0-1 
principle we can reduce the number of rounds of ShearSort to just one, resulting in overall 
merging time $T_m(M) = \OhOfV{\hat{t}\left(\frac{M}{w}\right)} = \OhOf{w\log^2 w}$.


\myparagraph{Distribution.}
Each merge round reduces the number of sequences $s$ by a factor of $2$, causing the 
algorithm to terminate after $\OhOfV{\log\left(\frac{N}{M}\right)}$ rounds. However, if $s<2P$, 
some CTAs will be idle while others have to work on longer sequences in each 
round. To improve resource utilization we suspend the merge at $s=P$ and distribute 
the sequences across $P$ buckets. To determine the bucket boundaries deterministically, 
we adapt the approach of Aggarwal and Vitter~\cite{aggarwal:io-model}. 

Given $s$ sorted sequences, each of size $\frac{N}{s}$, we pick $t$ evenly spaced elements 
as potential splitters from each sequence and sort the $st$ elements (let's call them~$S$). 
We pick $P-1$ evenly spaced items from $S$ as the final splitters that define bucket 
boundaries. To perform the distribution of the $s$ sequences into the $P$ buckets defined by 
the $P-1$ splitters, we calculate the rank of each splitter for every sorted sequence by 
parallel scans of the data, and sort the subsequences by bucket and sequence number.

\begin{lemma}\label{lemma:bucket-sizes}
Assuming all elements are distinct, the size of each bucket is at most 
$\OhOfV{\frac{N}{t} \cdot\left(1+\frac{t}{P}\right)}$ elements.
\end{lemma}

\begin{proof}
The number of elements within each sequence between any pair of potential splitters is 
$\ceilV{\frac{N}{st}}$. There are $\ceilV{\frac{st}{P}}$ potential splitters between a pair of 
selected final splitters. Thus, the total number of elements between final selected splitters 
is at most $s\cdot \ceilV{\frac{N}{st}} + \ceilV{\frac{st}{P}}\cdot\ceilV{\frac{N}{st}} = 
\OhOfV{\frac{N}{t}\cdot\left(1+\frac{t}{P}\right)}$.
\end{proof}

Thus, if we pick $t=P$ potential splitters from each sequence, the size of each bucket is at 
most $\OhOf{\frac{N}{P}}$. 

\myparagraph{Final merge phase.}
After rearranging the data the number of sorted subsequences is $P^2$. CTA $p_i$ 
then sorts the $i$th bucket by merging the sorted subsequences of the bucket. Since the subsequences may have 
sizes that are not a multiple of the page size, padding is introduced implicitly 
in shared memory, but not written to the output array. 

This phase requires $\log(P)$ additional merge rounds in which none of the $P$ CTAs 
runs out of work prematurely.

\begin{lemma}\label{lemma:bcfmergesort-runtime}
BCFMergesort takes $\OhOfV{\frac{N}{Pw}\cdot\left(1+\log^2(w)\cdot\log\left(\frac{N}{w}\right)\right)}$ 
time and requires $\OhOfV{1+\log\left(\frac{N}{M}\right)}$ parallel scans of the data in global memory.
\end{lemma}

\begin{proof}
The proof for Lemma~\ref{lemma:bcfmergesort-runtime} can be found in Appendix~\ref{sec:proof-lemma-2}.
\end{proof}

\subsection{Optimizations}
\label{sec:optimizations}
\myparagraph{Use of registers.} While latency of accessing shared memory is much lower than 
accessing global memory, it still incurs some latency. Any computation that requires 
multiple accesses to the same location in shared memory can be sped up by implementing it in 
registers. Since our GPU hardware has 64 physical registers per thread (63 of which can be 
used for data), and our ShearSort sorts $M = w^2$ elements, where $w = 32$, the sorting of each 
row (and column after the matrix transposition) of the $w \times w$ matrix can be 
implemented in registers. Note that the indices for accessing registers when loading
a row of $w=32$ elements do not depend on the thread index and the data is not 
transferred to local memory. The effects of register optimization are presented in Appendix~\ref{sec:register-optimization}.

\myparagraph{Implicit matrix transposition.} Processing the columns of the $w \times w$ 
matrix requires transposition of the matrix. The challenge with loading the column directly 
into registers is the fact that it either causes bank conflicts, or the indices of registers 
being loaded must depend on the thread identifier. The latter case causes the local memory  
(instead of actual registers) being used, which is actually implemented in off-chip global 
memory (see Section~\ref{sec:intro:orga}). Thus, neither of the approaches is acceptable. 
However, if we define column $j$ as {\em shifted diagonals} of the matrix (i.e. matrix 
elements $\M[(j+k) \mod w, k]$ for $0 \le k < w$)
and row $i$ to be the $i$th column, shifted upwards by $i$ elements (i.e. elements 
$\M[(i+k)\mod w,i]$ for $0 \le k < w$),
we can load each row and column directly into registers 
without the need for transposition.

\myparagraph{Duplicate keys.} 
Lemma~\ref{lemma:bucket-sizes} guarantees that bucket sizes are at most a constant factor 
more than $\frac{N}{P}$ only if all the items are distinct. A large number of duplicates 
might cause some bucket to be much larger than others. 
To mitigate this issue, we 
%
%
define additional $P-1$ buckets and place all items equal to the bucket boundary 
into their own bucket.  Thus, after the distribution, any duplicate items equal to 
bucket boundaries will already be in the correct position and do not need to be 
processed any further. We call this version {\em BCFMergesortPB}.

%% file: experiments.tex
\section{Experimental results}
\label{sec:experiments}
We conduct our experiments on a standard PC with a 2.4~GHz Intel Core2 Quad CPU, 
8~GB of RAM and an NVIDIA Geforce GTX580 GPU with 3~GB of memory, $P_h = 16$ SMs 
and $M_h = 48$KB per SM (with additional 16KB used for caching local memory). 
There are $w = 32$ physical cores per SM, $w = 32$ threads per warp and the 
shared memory consists of $w = 32$ banks. 
Our algorithms are implemented using the CUDA runtime API and compiled 
with the CUDA toolkit version~5.5, gcc version~4.7 and the -O~3 optimization flag.
The runtimes reported do not include the time needed to transfer data from the 
host computer to the GPU or back.
We compare our sorting algorithm to the well-known implementation of the Thrust 
library~\cite{bell:thrust} (version $1.7$) and warpsort~\cite{ye:sort}.\footnote{The results for warpsort presented in~\cite{ye:sort} included heuristics not mentioned in the paper. Unfortunately, these heuristics did not guarantee the runtimes mentioned in the paper and because of this the warpsort would run much slower for some inputs. Therefore, (with the approval of the authors of~\cite{ye:sort}) we modified warpsort to align with the description within their paper, resulting in more uniform runtimes, albeit slightly slower than what is presented in~\cite{ye:sort}.} We did not 
include CUDPP mergesort~\cite{davidson2012efficient} in our experiments, as the 
available code imposes severe limitations on the input and does not sort data larger 
than $64$~MB. Unfortunately, we also couldn't find a working implementation of the
Samplesort~\cite{gpu-sample-sort} for our hardware.

We report performance on four different data sets: {\em Random} data set consists of keys 
chosen uniformly from $[0\dots 10^6]$,  {\em distinct} data set is the sequence $1\dots n$
permuted by a shuffle algorithm~\cite{knuth:seminumerical}, {\em 0-1} data set contains 
uniformly chosen entries of zero or one and {\em defined duplicates} data set is formed by 
$k$ copies of each of $\frac{n}{k}$ distinct keys, randomly permuted by the shuffle algorithm. 


Warpsort uses a bucket distribution as well, but the splitter selection is randomized. 
It is known that deterministic algorithms result in significantly slower implementations in practice. Therefore, it is notable that our deterministic algorithms are only slightly slower than the randomized warpsort.

\begin{figure*}[tb]
\subfloat[]{
  \includegraphics[width=1.65in, angle=270]{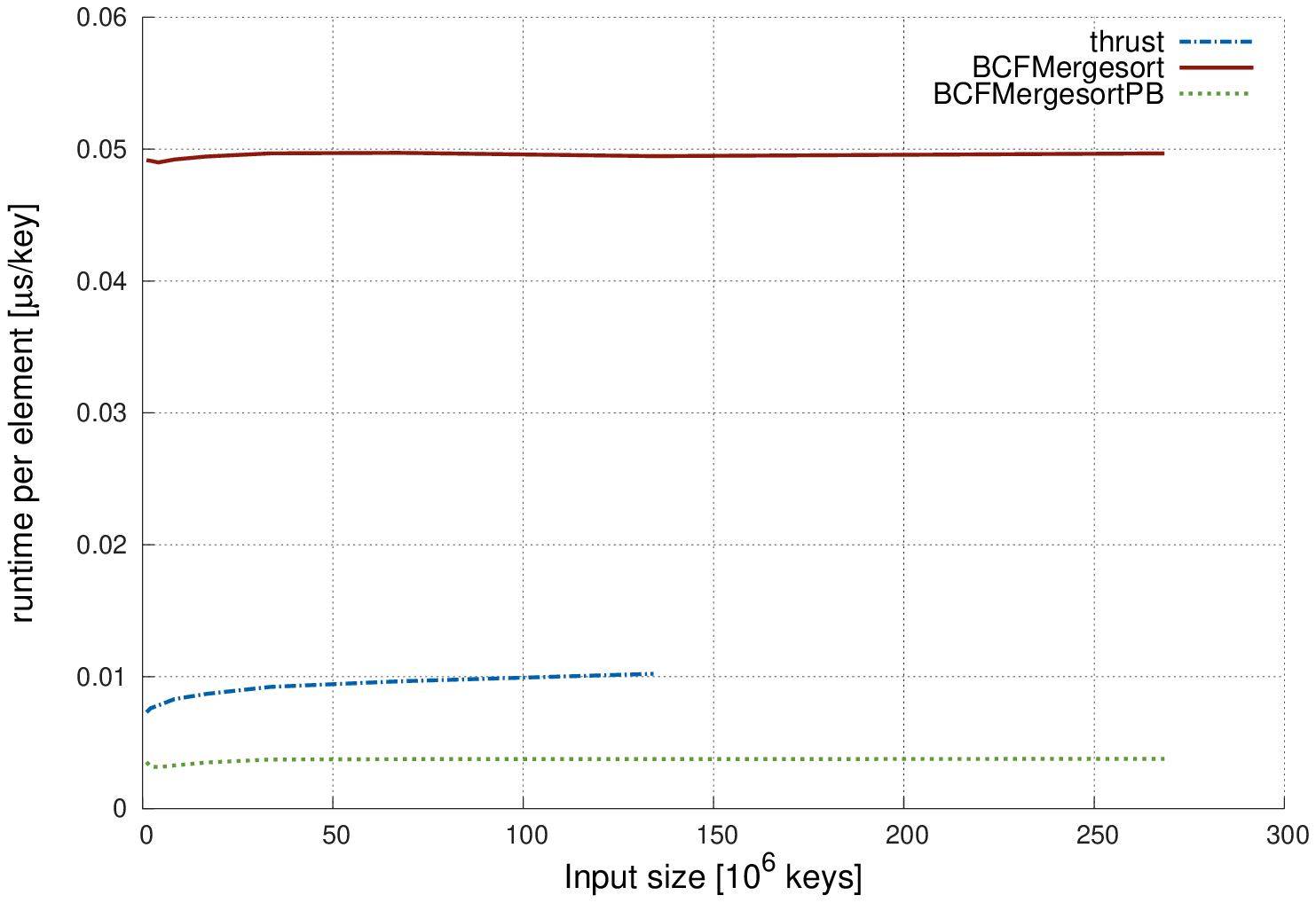}
  \label{fig:sort-throughput-01-defined-a}
}
~
\subfloat[]{
  \includegraphics[width=1.65in, angle=270]{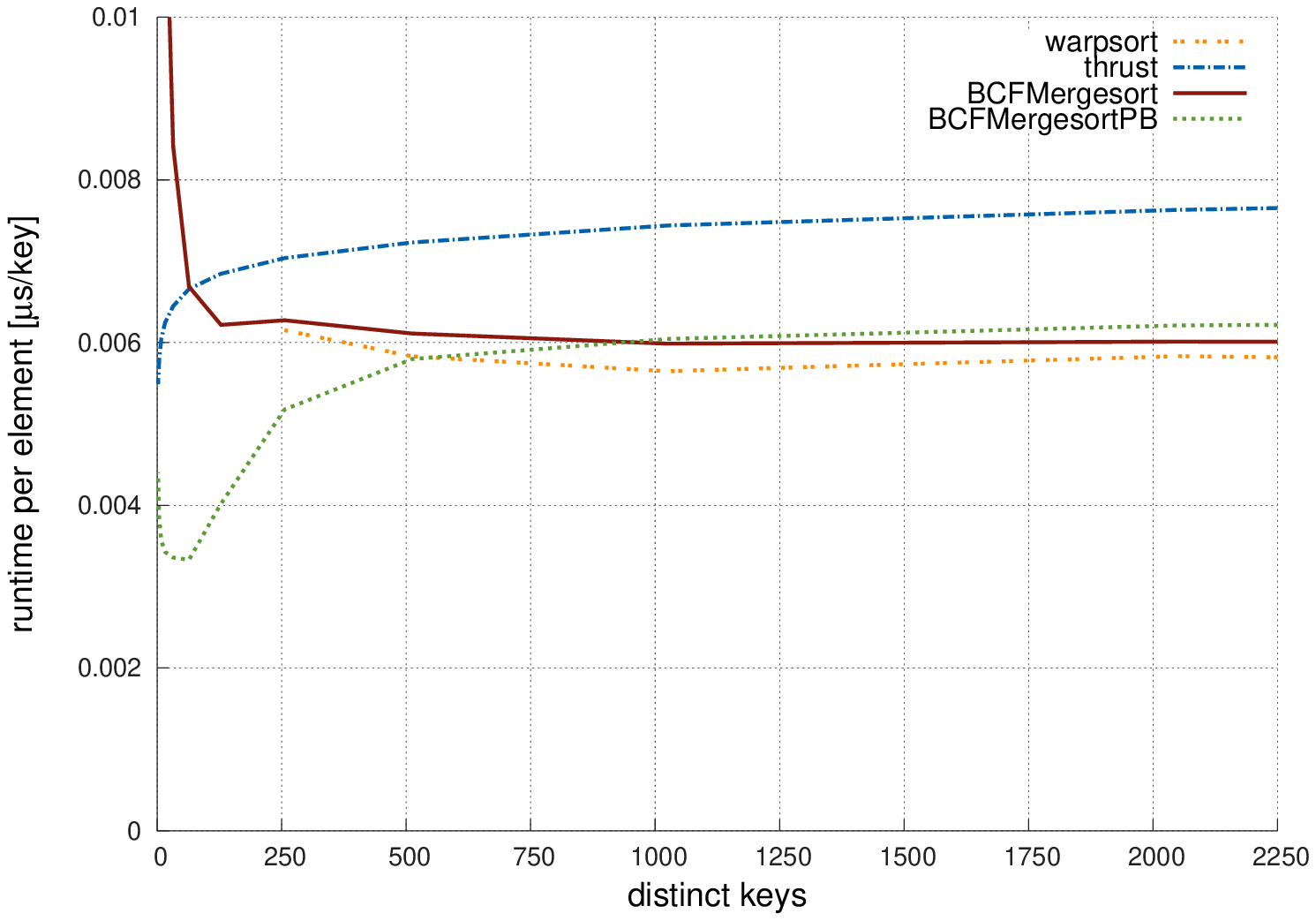}
  \label{fig:sort-throughput-01-defined-b}
}
\caption{Runtime per key for the {\em 0-1} (a) and 
	{\em defined duplicates} (b) data sets. For {\em defined duplicates} we limit the 
	axes to display the interesting part. BCFMergesort for two distinct keys takes 
	$0.051$ $\mu{}s$ per key. Input size for (b) is $2^{25}$ keys.}
\label{fig:sort-throughput-01-defined}
\end{figure*}

Duplicate keys skew the bucket sizes of the BCFMergesort. The effects of having separate buckets for pivot values can be seen in Figure~\ref{fig:sort-throughput-01-defined}. It's notable that warpsort fails to sort {\em 0-1} data completely. 

Though our merge algorithm is not in-place either, it requires less extra space than both Thrust and warpsort. BCFMergesort  
lets us sort up to 1GiB of data on our hardware, while Thrust and warpsort run on inputs up to 512 MiB and 256MiB, respectively.

The relevant parameters for the merge phase were empirically determined to be $p = 8P_h = 128$ and $M = w^2 = 1024$.



\begin{figure*}[tb]
\subfloat[]{
  \includegraphics[width=1.65in, angle=270]{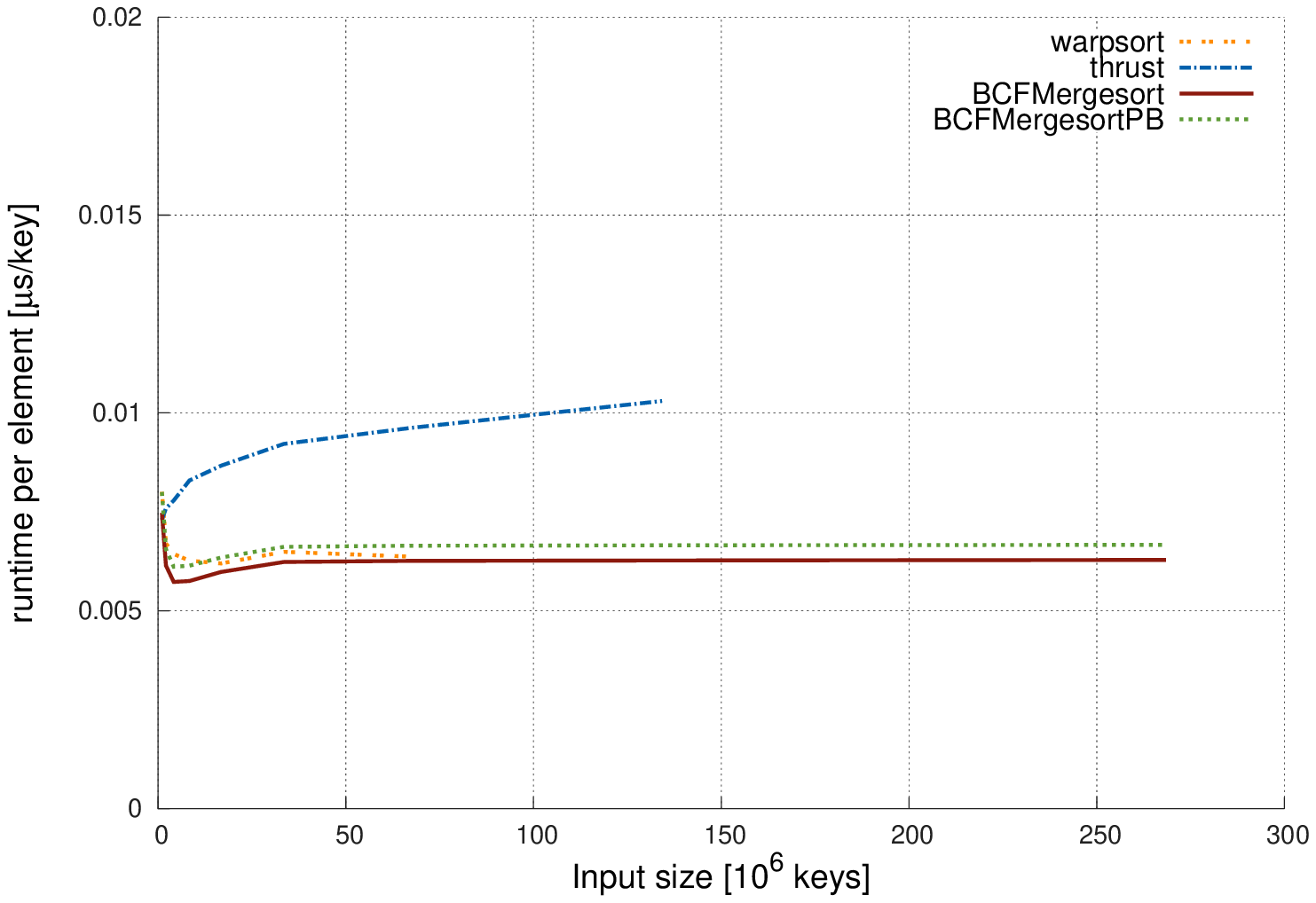}
  \label{fig:sort-throughput-all-a}
}
~
\subfloat[]{
  \includegraphics[width=1.65in, angle=270]{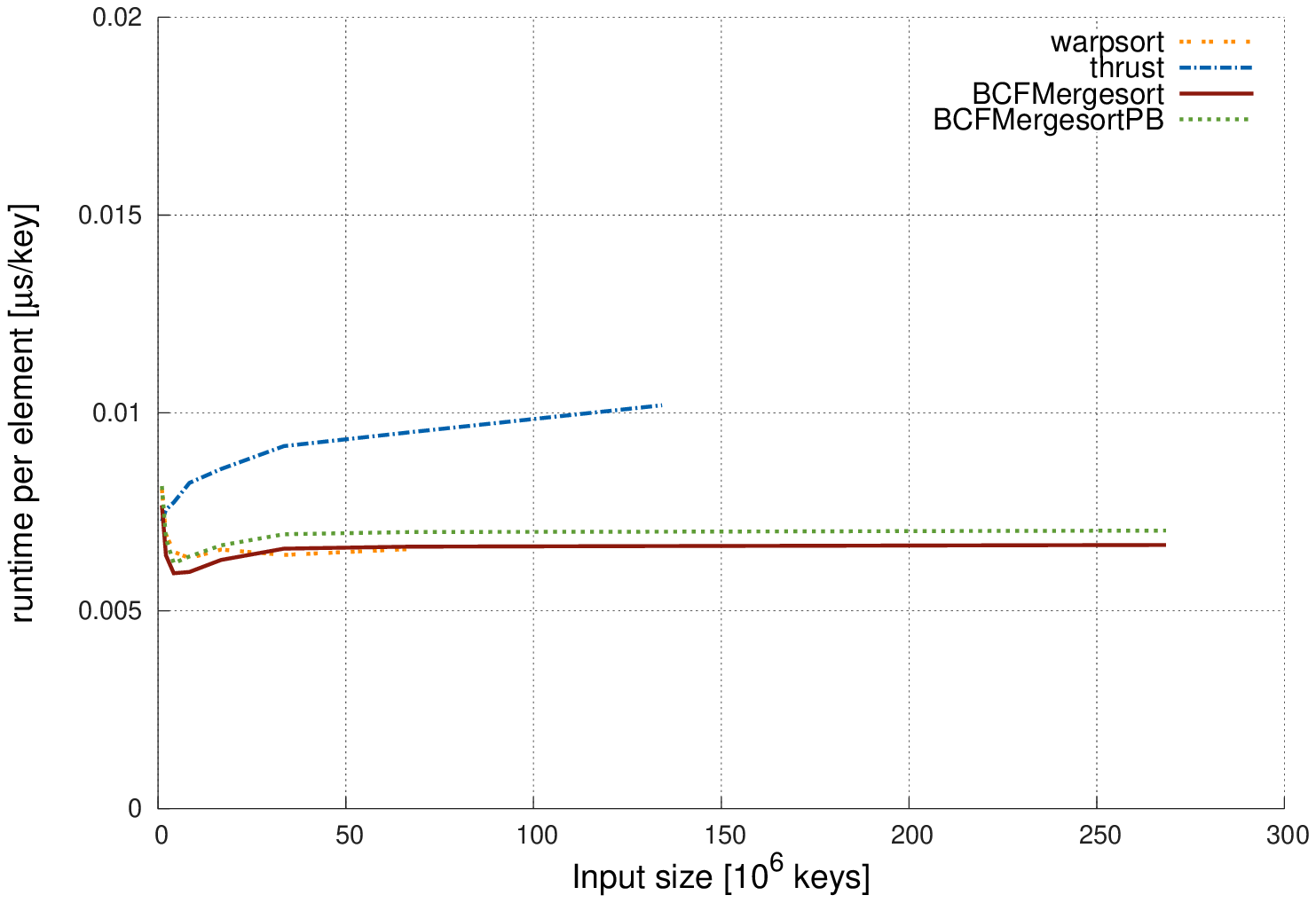}
  \label{fig:sort-throughput-all-b}
}
\caption{Runtime per key for the {\em random} (a) and {\em distinct} (b) data sets.}
\label{fig:sort-throughput-all}
\end{figure*}


The runtimes per key for the different algorithms and different input sets are shown in 
Figures~\ref{fig:sort-throughput-all} and~\ref{fig:sort-throughput-01-defined}. BCFMergesort
with pivot buckets (BCFMergesortPB) for inputs with few distinct keys suffers a performance loss of $6\%$ 
compared to our normal mergesort. While warpsort falls exactly into this gap for random keys,
the randomized pivot selection appears to be beneficial when dealing with inputs with all 
distinct keys, putting it on par and even surpassing our normal mergesort on two instances.
And while thrust mergesort is slightly faster than BCFMergesort on the smallest instance 
($5\%$), we beat thrust by up to $64\%$ on {\em random} inputs and $54\%$ on {\em distinct} 
keys.

The chart for {\em 0-1} data impressively shows the strength of pivot buckets introduced
at the end of Section~\ref{sec:merging}. Since the heuristic causes BCFMergesortPB to have 
nothing more to do after rearranging the data, BCFMergesortPB beats 
BCFMergesort on these inputs by a factor of up to $15.5$ and thrust sort by a factor of up 
to $2.7$. In the chart for {\em defined duplicates}, the less extreme versions of up to 
$2250$ distinct keys can be seen. BCFMergesortPB first surpasses the normal version of 
BCFMergesort when $\frac{1}{4}$ of all keys are equal to a splitter.

%% file: conclusion.tex
\section{Conclusion}
%
We discussed the properties of GPUs, focusing on the memory hierarchy. We 
provide experimental data that shows the great impact of shared memory 
bank conflicts on the runtime of GPU programs and described methods to design
bank conflict free algorithms for GPUs. Using those methods, we developed 
the first comparison-based bank conflict free sorting algorithm for GPUs and showed that the
benefits of having no bank conflicts outweigh the additional work necessary
to make algorithms bank conflict free.


\myparagraph{Acknowledgments.}
We would like to thank Vitaly Osipov for helpful discussions and sharing his
insights on GPUs.

%% file: prefix_sums.tex
\section{Impact of Bank Conflicts on Runtimes} 
\label{sec:colored-prefix-sums}
The importance of shared memory bank conflicts can easily be demonstrated by looking at the 
problem of {\em colored prefix sums}: given an array of $N$ elements and a set of $d$ colors 
$\{c_1, c_2, \dots, c_d\}$, with a color $c_i$ associated with each element, colored prefix 
sums asks to compute $d$ independent prefix sums among the elements of the same color.

Our algorithm is basically a three-phase prefix sums algorithm similar to 
Merrill's~\cite{merrill:scan}. The elements are assigned to $P$ CTAs in contiguous 
tiles of size $\frac{N}{P}$. For colored prefix sums, we keep intermediate values for each 
color and perform extra operations to identify the color of each element. 

In the first phase all elements of the tiles are reduced using the prefix sums operation. 
This results in $d$ values per CTA, one for each color. The second phase calculates 
the prefix sums of those $P\cdot d$ reduced values as offsets for the neighboring tiles. 
This can be done on a single CTA interpreting the values as a $d\times P$ matrix.
In the final phase, the prefix sums of the tiles are calculated, seeded with the offsets 
computed in the second phase. Since the tiles may be too large for shared memory, they are 
loaded in contiguous pages of $M$ elements. Between neighboring pages, page offsets for each 
color have to be stored in shared memory. The calculation of colored prefix sums within a 
page resembles the algorithm itself: Each thread is assigned a contiguous chunk of the page
and first counts the entries per color in its chunk, then the intra-page offsets are 
calculated by one scan of those sums per color, and finally the threads calculate colored 
prefix sums on their chunks seeded with the tile-, inter-page and intra-page offsets. 


We would expect the runtime to depend on the input size $N$, the number of CTAs 
$P$ and the warp size $w$, but not $d$ since all calculations depending on the number of 
colors are performed in parallel. However, the runtime increases with the number of colors 
(see Figure~\ref{fig:cps-runtime}). The slowdown correlates nicely with the number of shared 
memory bank conflicts incurred when calculating intra-page offsets in the third phase.

To improve our algorithm, we implement a way to calculate the intra-page offsets without any 
bank conflicts. We assign each thread one memory bank for storage of intermediate values. 
The resulting $w \times d$ matrix lists the color values in columns. To avoid having 
multiple threads working on the same banks we transpose the matrix and then calculate the 
prefix sums with $d$ threads in parallel. After retransposing the matrix, we can propagate 
the offsets.

We observe that the runtime of the algorithm is now constant for different numbers of colors 
(see Figure~\ref{fig:cps-runtime}). We limit the number of colors to $16$ because of the 
limited amount of shared memory available for each multiprocessor. 

%% file: proof-lemma-2.tex
\section{Proof for Lemma~\ref{lemma:bcfmergesort-runtime}}
\label{sec:proof-lemma-2}
\begin{proof}
The BCFMergesort runs in four basic phases: the base case phase, the first merge 
phase, the bucket distribution and the final merge phase. We will look at each 
phase separately and combine the results later.

The base case phase uses $P$ CTAs working in parallel to sort the input in 
$\frac{N}{M}$ segments of size $M$. Each CTA has to perform 
$\ceilV{\frac{N}{M\cdot P}}$ rounds of sorting, resulting in 
$\OhOfV{\frac{N}{M\cdot P}\cdot \hat{t}(w) \cdot \log(w)}$ parallel time. In the 
process, each data item is read and written once.

To reduce the number of sorted sequences to $P$, 
$\log\left(\frac{N}{M\cdot P}\right)$ rounds of binary merge have to be 
performed in the first merge phase. To merge two sequences of size $\frac{N}{2}$, 
a total of $2\cdot \frac{N/2}{M/2} = \frac{2\cdot N}{M}$ pages of size 
$\frac{M}{2}$ have to be processed in each round: one page is read, two pages 
are merged, one page is written. Therefore, each element is merged twice in 
average, resulting in a parallel time of 
$\OhOfV{\frac{N}{M\cdot P}\cdot\log\left(\frac{N}{M\cdot P}\right)\cdot\hat{t}(w)}$
and $\log\left(\frac{N}{M\cdot P}\right)$ additional reads and writes of the data. 

The bucket distribution in itself consists of five phases: picking $t$ pivots 
from each sorted sequence, sorting those $t\cdot P$ pivots, choosing the $P-1$
final pivots and finding their location in the sorted sequences, calculating the
splitter offsets and rearranging the data by bucket. Since we use one CTA 
per sorted sequence, selecting $t$ pivots takes $\OhOfV{\frac{t}{w}}$ time. 

The pivots are then sorted using mergesort: base case and binary merge phases. 
That results in $\OhOfV{\frac{P^2}{M}\cdot\hat{t}(w)\cdot\log(w)+
\log\left(\frac{P^2}{M}\right)\cdot\hat{t}(w)} 
= \OhOfV{\frac{P^2}{M}\cdot\hat{t}(w)\cdot\log(w)}$ parallel time. 

The splitter are located by sequentially scanning the sequences using $w$ 
threads for each. Therefore, each CTA needs $\OhOfV{\frac{N}{Pw}}$ time.
In the process, the data is read once.

The calculation of the $P^2$ subsequence offsets takes another $\OhOf{P}$ time.

Finally, the data is rearranged. Each of the $P$ CTAs moves $P$ 
subsequences to their proper positions, causing $\OhOf{P}$ computation and
one more read and write of all data.

The final merge phase causes another $\log(P)$ binary merges. Implicit padding 
can at most be $\frac{M}{2}-1$ elements per subsequence, 
$\left(\frac{M}{2}-1\right)\cdot P^2$ in total. This is $\OhOf{N}$ for any 
reasonable combination of parameters, so the time of the page-wise binary merge 
applies.

This results in a total parallel runtime of 
\begin{eqnarray*}
& & \OhOfV{\frac{N}{M\cdot P}\cdot\hat{t}(w)\cdot \log(w)} + \OhOfV{\frac{N}{M\cdot P}\cdot\log\left(\frac{N}{M\cdot P}\right)\cdot\hat{t}(w)} \\
&+& \OhOfV{\frac{t}{w}} + \OhOfV{\frac{P^2}{M}\cdot\hat{t}(w)\cdot\log(w)} + \OhOfV{\frac{N}{Pw}} + \OhOf{P} +\OhOf{P} \\
&+& \OhOfV{\frac{N}{M\cdot P}\cdot\log(P)\cdot\hat{t}(w)} \\
&=& \OhOfV{\frac{N}{M\cdot P}\cdot\hat{t}(w)\cdot \log(w)+\frac{N}{M\cdot P}\cdot\log\left(\frac{N}{M\cdot P}\right)\cdot\hat{t}(w)+ \frac{N}{Pw} + \frac{N}{M\cdot P}\cdot\log(P)\cdot\hat{t}(w)} \\
&=& \OhOfV{\frac{N}{M\cdot P}\cdot\hat{t}(w)\cdot\left(\log(w)+\log\left(\frac{N}{M\cdot P}\right)+\log(P)\right) + \frac{N}{Pw}} \\
&=& \OhOfV{\frac{N}{M\cdot P}\cdot\hat{t}(w)\cdot\log\left(\frac{N\cdot w}{M}\right) + \frac{N}{Pw}}. \\ \\
& & \text{Considering that $M=w^2$ and $\hat{t}(w)=w\cdot\log^2(w)$, we get } \\ \\
& & \OhOfV{\frac{N}{M\cdot P}\cdot\hat{t}(w)\cdot\log\left(\frac{N\cdot w}{M}\right) + \frac{N}{Pw}} \\
&=& \OhOfV{\frac{N}{Pw}\cdot\log^2(w)\cdot\log\left(\frac{N}{w}\right) + \frac{N}{Pw}} \\
&=& \OhOfV{\frac{N}{Pw}\cdot\left(1+\log^2(w)\cdot\log\left(\frac{N}{w}\right)\right)}.
\end{eqnarray*}

and in $\OhOfV{1+\log\left(\frac{N}{M\cdot P}\right)+\log(P)} = \OhOfV{1+\log\left(\frac{N}{M}\right)}$
passes over the data.
\end{proof}

%% file: additional-charts.tex
\section{Additional charts}
\label{sec:additional-charts}

\begin{figure*}[bh]
\subfloat[]{
  \includegraphics[width=1.65in, angle=270]{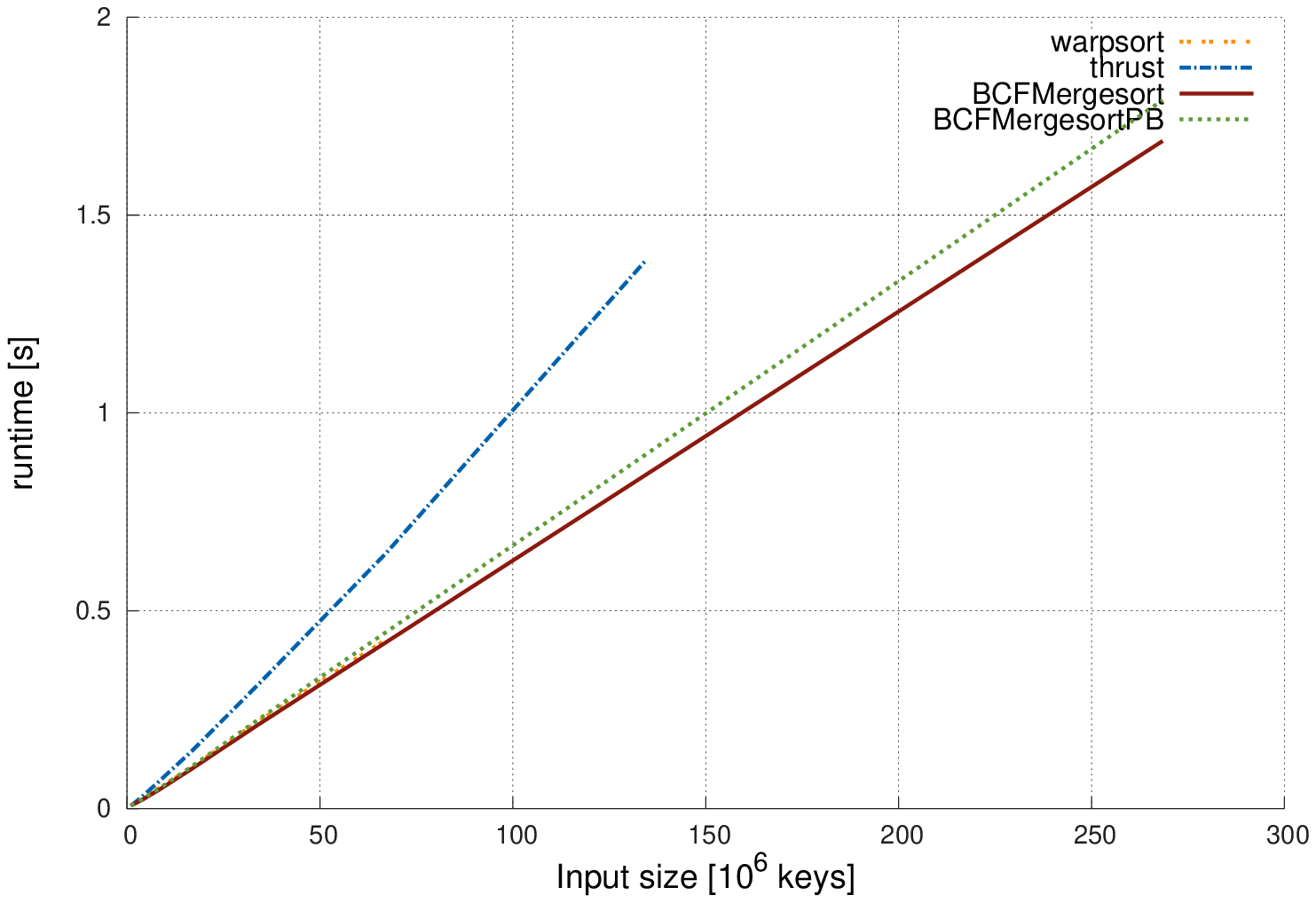}
  \label{fig:sort-runtime-all-a}
}
~
\subfloat[]{
  \includegraphics[width=1.65in, angle=270]{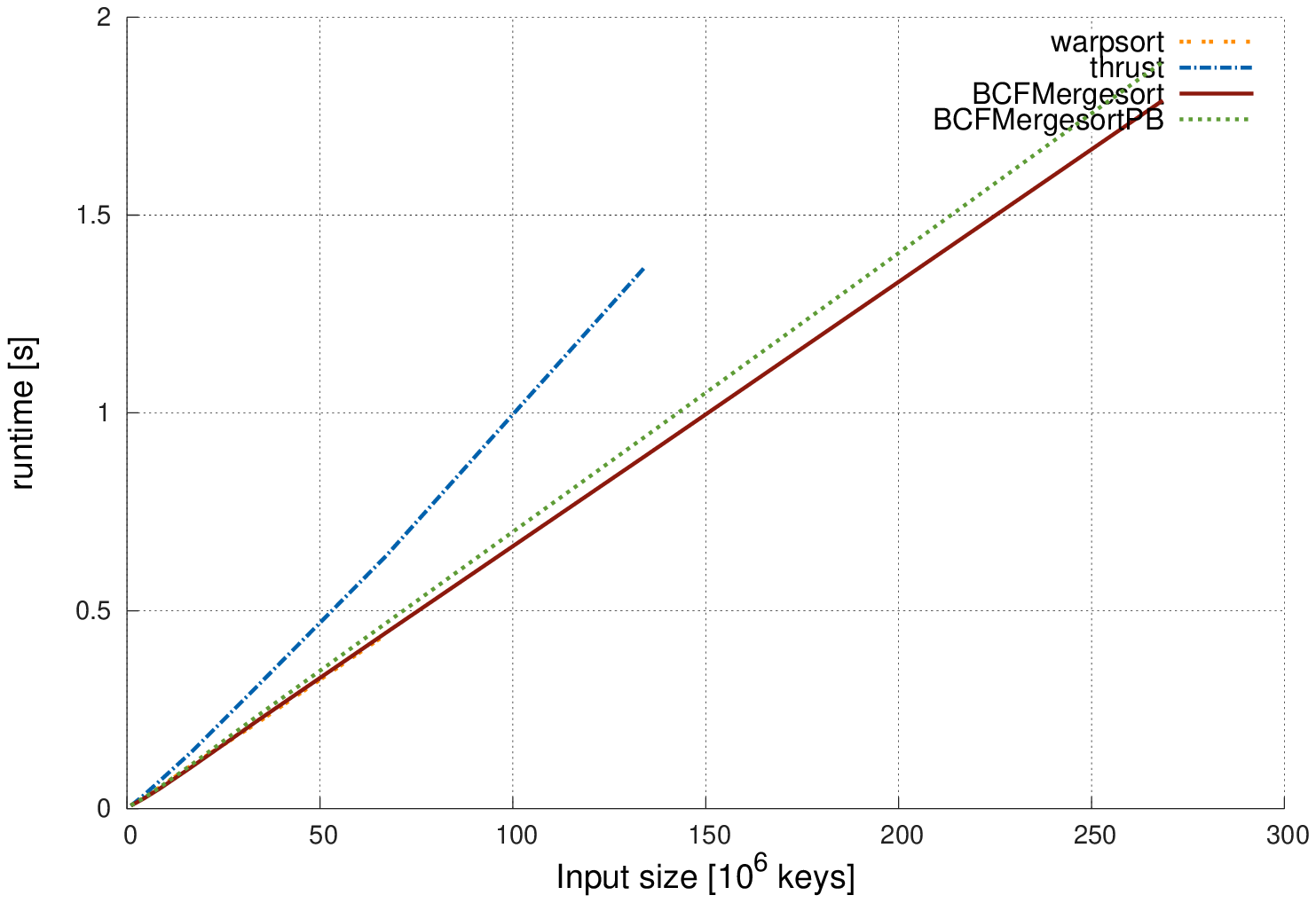}
  \label{fig:sort-runtime-all-b}
}
\caption{Runtime for the {\em random} (a) and {\em distinct} (b) data sets.}
\label{fig:sort-runtime-all}
\end{figure*}

\begin{figure*}[bh]
\subfloat[]{
  \includegraphics[width=1.65in, angle=270]{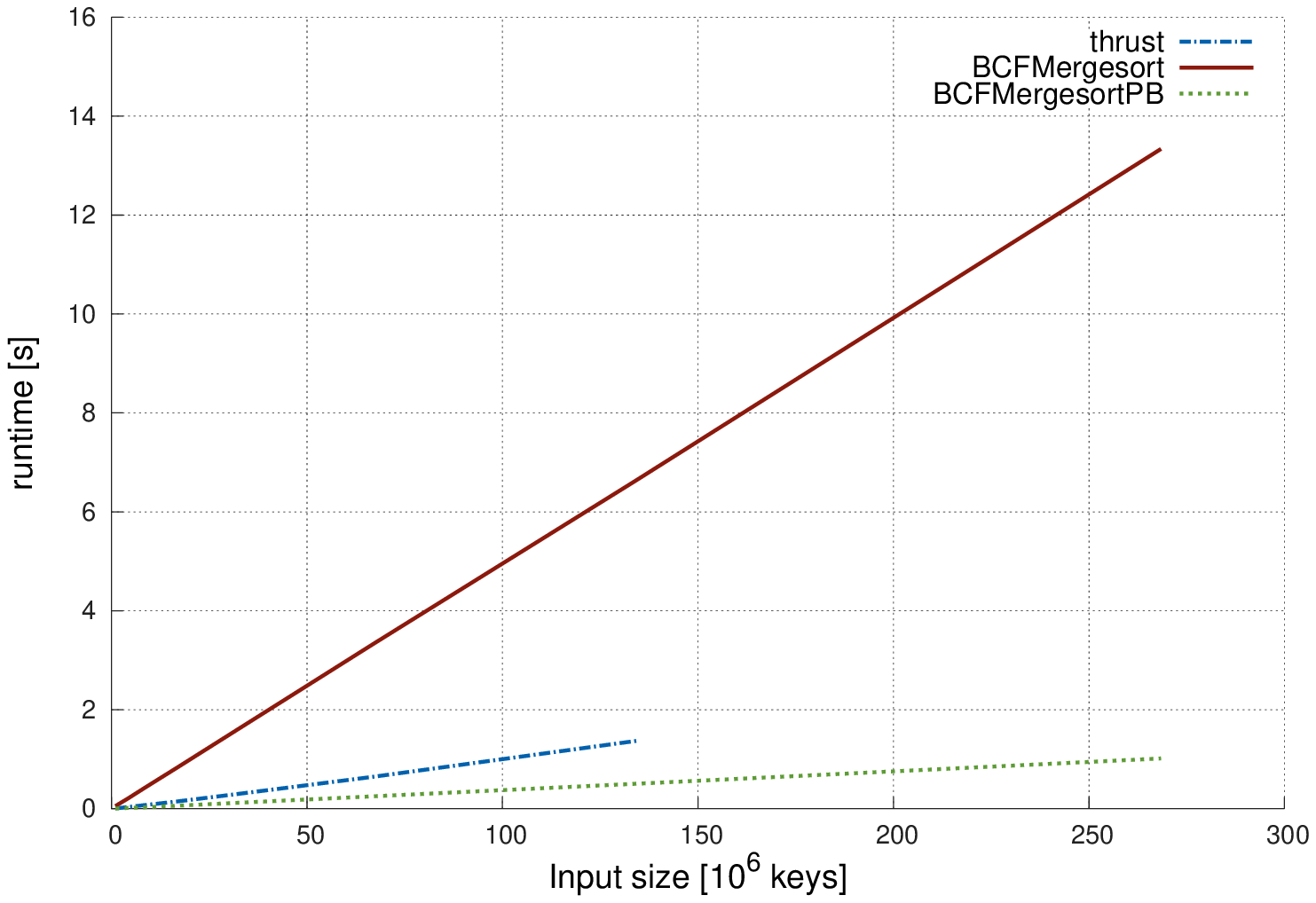}
  \label{fig:sort-runtime-01-defined-a}
}
~
\subfloat[]{
  \includegraphics[width=1.65in, angle=270]{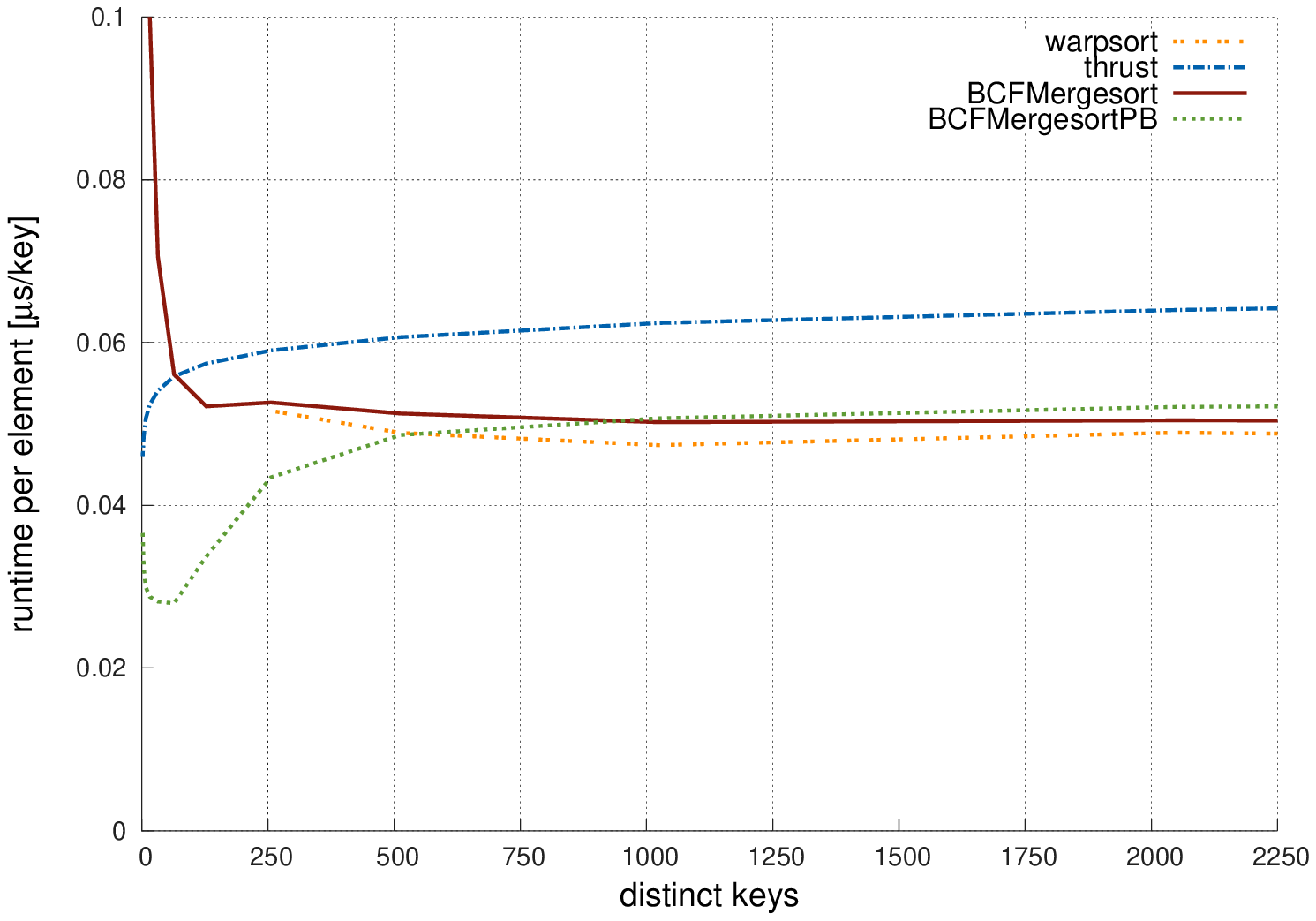}
  \label{fig:sort-runtime-01-defined-b}
}
\caption{Runtimes for the {\em 0-1} (a) and {\em defined duplicates} (b) data sets. For 
	{\em defined duplicates} we limit the axes to display the interesting part. BCFMergesort 
	for two distinct keys takes $0.051$ $\mu{}s$ per key. Input size for (b) is $2^{25}$ keys.}
\label{fig:sort-runtime-01-defined}
\end{figure*}

For completeness we present the plain runtime charts of GPU mergesort algorithms 
in Figures~\ref{fig:sort-runtime-all} and~\ref{fig:sort-runtime-01-defined}. The dominance of the linear term makes it difficult to see the lower order terms, for sorting, which is the reason why we chose to normalize the runtimes per element in the main presentation of the results.


%

%% file: register-optimization.tex
\section{Register optimization}
\label{sec:register-optimization}
\begin{figure*}[bt]
	\begin{center}
	\includegraphics[width=2in, angle=270]{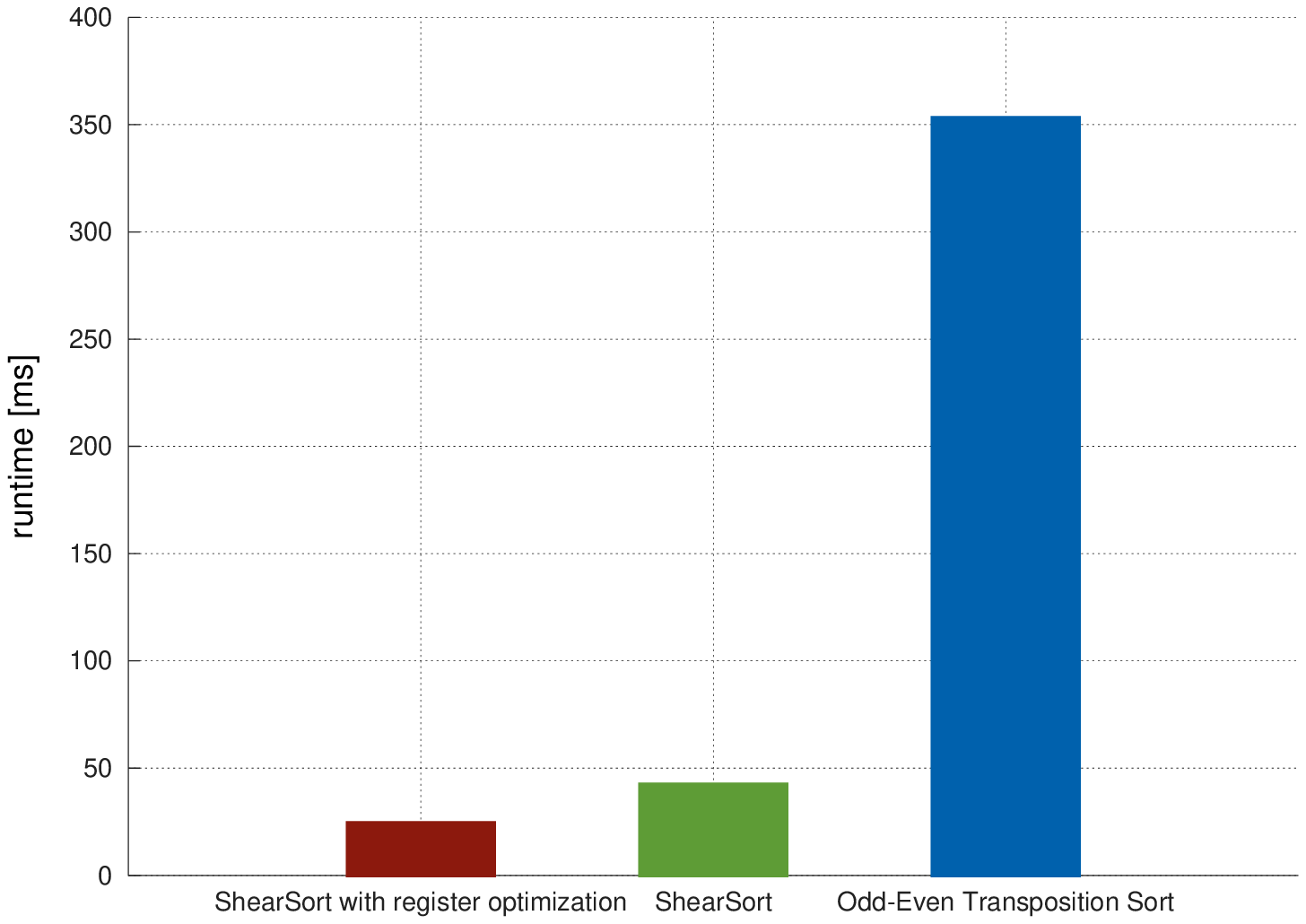}
	\end{center}
	\caption{Comparison of runtimes of ShearSort with and without register optimization and of Odd-Even Transposition sorting network on subsets of 1024 elements each, for the total of size of $2^{25}$ elements.}
	\label{fig:sort-optimization}
\end{figure*}

We ran experiments to determine the speedup provided by sorting rows and columns 
in register space. The results in Figure~\ref{fig:sort-optimization} show the runtimes
of BCFMergesort with transposition on a $w\times w$ matrix with register optimization
and without. As a comparison, we also provide the runtime of an implementation of 
Odd-Even Transposition Sort on $w^2 = 1024$ elements. It demonstrates the
effect of bank conflicts in shared memory on runtimes. The sorting algorithms are applied to subsets of $w^2$ elements on the total input of $2^{25}$ elements.

%% file: shearsort.tex
\section{Bank-conflict free computation on non-square matrices}
\label{sec:long-shearsort}
The approach to bank conflict free computation from Section~\ref{sec:bank-conflicts} 
can easily be generalized for inputs of size $n = hw^2$ for a small
integer constant $h$ that divides $w$, by transforming the input from 
column-major to row-major order and vice versa. Since space in shared memory 
is limited~\cite{cudaprogrammingguide}, it is desirable to perform this 
transformation in-place. Although there is no simple in-place algorithm for 
transforming general rectangular matrices from row-major to column-major 
order, $h$ is a sufficiently small integer for current sizes of shared 
memories on GPUs (currently $h \le 12$). 

\begin{figure*}[th]
	\begin{center}
		\begin{tikzpicture}[scale=.13,auto,every node/.style={inner sep=0pt,outer sep=0pt,circle,minimum size=10pt,font=\tiny}]
		\def\horizstart{1}
		\def\horizspace{2}
		\def\vertstart{1}
		\def\vertspace{2}
		\def\labelstart{60}
		\definecolor{procolor}{RGB}{0,255,255}
		\useasboundingbox (\horizstart+0*\horizspace,\vertstart) rectangle (\horizstart+32*\horizspace,\vertstart+8*\vertspace);
		\draw[black,fill=green]   (\horizstart+0*\horizspace,\vertstart+7*\vertspace) rectangle (\horizstart+32*\horizspace,\vertstart+8*\vertspace);
		\draw[black,fill=yellow]  (\horizstart+0*\horizspace,\vertstart+6*\vertspace) rectangle (\horizstart+32*\horizspace,\vertstart+7*\vertspace);
		\draw[black,fill=white]   (\horizstart+0*\horizspace,\vertstart+5*\vertspace) rectangle (\horizstart+32*\horizspace,\vertstart+6*\vertspace);
		\draw[black,fill=blue]    (\horizstart+0*\horizspace,\vertstart+4*\vertspace) rectangle (\horizstart+32*\horizspace,\vertstart+5*\vertspace);
		\draw[black,fill=black]   (\horizstart+0*\horizspace,\vertstart+3*\vertspace) rectangle (\horizstart+32*\horizspace,\vertstart+4*\vertspace);
		\draw[black,fill=cyan]    (\horizstart+0*\horizspace,\vertstart+2*\vertspace) rectangle (\horizstart+32*\horizspace,\vertstart+3*\vertspace);
		\draw[black,fill=magenta] (\horizstart+0*\horizspace,\vertstart+1*\vertspace) rectangle (\horizstart+32*\horizspace,\vertstart+2*\vertspace);
		\draw[black,fill=red]     (\horizstart+0*\horizspace,\vertstart+0*\vertspace) rectangle (\horizstart+32*\horizspace,\vertstart+1*\vertspace);
		\draw[dashed] (\horizstart+8*\horizspace,\vertstart+8.5*\vertspace) -> (\horizstart+8*\horizspace,\vertstart-0.5*\vertspace);
		\draw[dashed] (\horizstart+16*\horizspace,\vertstart+8.5*\vertspace) -> (\horizstart+16*\horizspace,\vertstart-0.5*\vertspace);
		\draw[dashed] (\horizstart+24*\horizspace,\vertstart+8.5*\vertspace) -> (\horizstart+24*\horizspace,\vertstart-0.5*\vertspace);
		\end{tikzpicture}
		
		\vspace{5mm}
		
		\begin{tikzpicture}[scale=.13,auto,every node/.style={inner sep=0pt,outer sep=0pt,circle,minimum size=10pt,font=\tiny}]
		\def\horizstart{1}
		\def\horizspace{2}
		\def\vertstart{1}
		\def\vertspace{2}
		\def\labelstart{60}
		\definecolor{procolor}{RGB}{0,255,255}
		\useasboundingbox (\horizstart+0*\horizspace,\vertstart) rectangle (\horizstart+32*\horizspace,\vertstart+8*\vertspace);
		\draw[black,fill=green]   (\horizstart+0*\horizspace,\vertstart+0*\vertspace) rectangle (\horizstart+1*\horizspace,\vertstart+8*\vertspace);
		\draw[black,fill=yellow]  (\horizstart+1*\horizspace,\vertstart+0*\vertspace) rectangle (\horizstart+2*\horizspace,\vertstart+8*\vertspace);
		\draw[black,fill=white]   (\horizstart+2*\horizspace,\vertstart+0*\vertspace) rectangle (\horizstart+3*\horizspace,\vertstart+8*\vertspace);
		\draw[black,fill=blue]    (\horizstart+3*\horizspace,\vertstart+0*\vertspace) rectangle (\horizstart+4*\horizspace,\vertstart+8*\vertspace);
		\draw[black,fill=black]   (\horizstart+4*\horizspace,\vertstart+0*\vertspace) rectangle (\horizstart+5*\horizspace,\vertstart+8*\vertspace);
		\draw[black,fill=cyan]    (\horizstart+5*\horizspace,\vertstart+0*\vertspace) rectangle (\horizstart+6*\horizspace,\vertstart+8*\vertspace);
		\draw[black,fill=magenta] (\horizstart+6*\horizspace,\vertstart+0*\vertspace) rectangle (\horizstart+7*\horizspace,\vertstart+8*\vertspace);
		\draw[black,fill=red]     (\horizstart+7*\horizspace,\vertstart+0*\vertspace) rectangle (\horizstart+8*\horizspace,\vertstart+8*\vertspace);
		\draw[black,fill=green]   (\horizstart+8*\horizspace,\vertstart+0*\vertspace) rectangle (\horizstart+9*\horizspace,\vertstart+8*\vertspace);
		\draw[black,fill=yellow]  (\horizstart+9*\horizspace,\vertstart+0*\vertspace) rectangle (\horizstart+10*\horizspace,\vertstart+8*\vertspace);
		\draw[black,fill=white]   (\horizstart+10*\horizspace,\vertstart+0*\vertspace) rectangle (\horizstart+11*\horizspace,\vertstart+8*\vertspace);
		\draw[black,fill=blue]    (\horizstart+11*\horizspace,\vertstart+0*\vertspace) rectangle (\horizstart+12*\horizspace,\vertstart+8*\vertspace);
		\draw[black,fill=black]   (\horizstart+12*\horizspace,\vertstart+0*\vertspace) rectangle (\horizstart+13*\horizspace,\vertstart+8*\vertspace);
		\draw[black,fill=cyan]    (\horizstart+13*\horizspace,\vertstart+0*\vertspace) rectangle (\horizstart+14*\horizspace,\vertstart+8*\vertspace);
		\draw[black,fill=magenta] (\horizstart+14*\horizspace,\vertstart+0*\vertspace) rectangle (\horizstart+15*\horizspace,\vertstart+8*\vertspace);
		\draw[black,fill=red]     (\horizstart+15*\horizspace,\vertstart+0*\vertspace) rectangle (\horizstart+16*\horizspace,\vertstart+8*\vertspace);
		\draw[black,fill=green]   (\horizstart+16*\horizspace,\vertstart+0*\vertspace) rectangle (\horizstart+17*\horizspace,\vertstart+8*\vertspace);
		\draw[black,fill=yellow]  (\horizstart+17*\horizspace,\vertstart+0*\vertspace) rectangle (\horizstart+18*\horizspace,\vertstart+8*\vertspace);
		\draw[black,fill=white]   (\horizstart+18*\horizspace,\vertstart+0*\vertspace) rectangle (\horizstart+19*\horizspace,\vertstart+8*\vertspace);
		\draw[black,fill=blue]    (\horizstart+19*\horizspace,\vertstart+0*\vertspace) rectangle (\horizstart+20*\horizspace,\vertstart+8*\vertspace);
		\draw[black,fill=black]   (\horizstart+20*\horizspace,\vertstart+0*\vertspace) rectangle (\horizstart+21*\horizspace,\vertstart+8*\vertspace);
		\draw[black,fill=cyan]    (\horizstart+21*\horizspace,\vertstart+0*\vertspace) rectangle (\horizstart+22*\horizspace,\vertstart+8*\vertspace);
		\draw[black,fill=magenta] (\horizstart+22*\horizspace,\vertstart+0*\vertspace) rectangle (\horizstart+23*\horizspace,\vertstart+8*\vertspace);
		\draw[black,fill=red]     (\horizstart+23*\horizspace,\vertstart+0*\vertspace) rectangle (\horizstart+24*\horizspace,\vertstart+8*\vertspace);
		\draw[black,fill=green]   (\horizstart+24*\horizspace,\vertstart+0*\vertspace) rectangle (\horizstart+25*\horizspace,\vertstart+8*\vertspace);
		\draw[black,fill=yellow]  (\horizstart+25*\horizspace,\vertstart+0*\vertspace) rectangle (\horizstart+26*\horizspace,\vertstart+8*\vertspace);
		\draw[black,fill=white]   (\horizstart+26*\horizspace,\vertstart+0*\vertspace) rectangle (\horizstart+27*\horizspace,\vertstart+8*\vertspace);
		\draw[black,fill=blue]    (\horizstart+27*\horizspace,\vertstart+0*\vertspace) rectangle (\horizstart+28*\horizspace,\vertstart+8*\vertspace);
		\draw[black,fill=black]   (\horizstart+28*\horizspace,\vertstart+0*\vertspace) rectangle (\horizstart+29*\horizspace,\vertstart+8*\vertspace);
		\draw[black,fill=cyan]    (\horizstart+29*\horizspace,\vertstart+0*\vertspace) rectangle (\horizstart+30*\horizspace,\vertstart+8*\vertspace);
		\draw[black,fill=magenta] (\horizstart+30*\horizspace,\vertstart+0*\vertspace) rectangle (\horizstart+31*\horizspace,\vertstart+8*\vertspace);
		\draw[black,fill=red]     (\horizstart+31*\horizspace,\vertstart+0*\vertspace) rectangle (\horizstart+32*\horizspace,\vertstart+8*\vertspace);
		\draw[dashed] (\horizstart+8*\horizspace,\vertstart+8.5*\vertspace) -> (\horizstart+8*\horizspace,\vertstart-0.5*\vertspace);
		\draw[dashed] (\horizstart+16*\horizspace,\vertstart+8.5*\vertspace) -> (\horizstart+16*\horizspace,\vertstart-0.5*\vertspace);
		\draw[dashed] (\horizstart+24*\horizspace,\vertstart+8.5*\vertspace) -> (\horizstart+24*\horizspace,\vertstart-0.5*\vertspace);
		\end{tikzpicture}
	\end{center}
	\caption{$8\times 32$ matrix with division into $8\times 8$-submatrices, before and after transforming the square submatrices}
	\label{fig:long-shearsort-1}
\end{figure*}
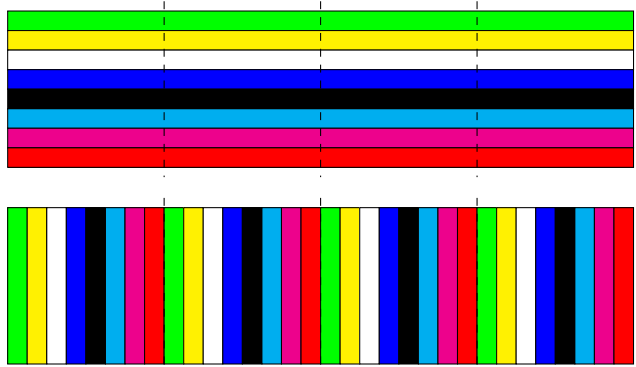

To transform the matrix from column-major to row-major layout, we split the $w
\times hw$ matrix into $h$ square $w \times w$ matrices, transpose each one
using a different warp (see Figure~\ref{fig:long-shearsort-1}), and combine 
$h$ original columns (now $h$ partial rows) that compose a single $hw$-sized row 
using all $h$ warps (see Figure~\ref{fig:long-shearsort-3}). Transformation from
row-major to column-major layout is done similarly.
For pseudocode refer to Appendix~\ref{sec:pseudocode}.


\vspace{5mm}

Then all $h$ columns resulting from the transformation of the same row are gathered 
together in the correct order. 
The gathering of all $h$ columns resulting from the transformation of the same 
row can be done by $h\cdot w$ threads in parallel without bank conflicts, 
because the $w$ threads of the same warp will always run in lock-step. Each 
thread is responsible for one element of each resulting column of $\frac{w}{h}$ rows. 

\vspace{5mm}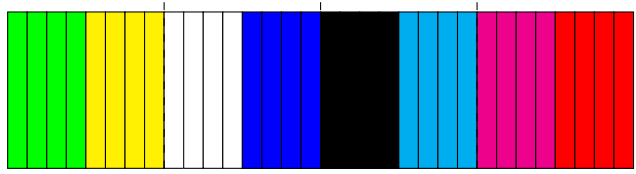
\begin{figure*}[th]
	\begin{center}
		\begin{tikzpicture}[scale=.13,auto,every node/.style={inner sep=0pt,outer sep=0pt,circle,minimum size=10pt,font=\tiny}]
		\def\horizstart{1}
		\def\horizspace{2}
		\def\vertstart{1}
		\def\vertspace{2}
		\def\labelstart{60}
		\definecolor{procolor}{RGB}{0,255,255}
		\useasboundingbox (\horizstart+0*\horizspace,\vertstart) rectangle (\horizstart+32*\horizspace,\vertstart+8*\vertspace);
		\draw[black,fill=green]   (\horizstart+0*\horizspace,\vertstart+0*\vertspace) rectangle (\horizstart+1*\horizspace,\vertstart+8*\vertspace);
		\draw[black,fill=green]  (\horizstart+1*\horizspace,\vertstart+0*\vertspace) rectangle (\horizstart+2*\horizspace,\vertstart+8*\vertspace);
		\draw[black,fill=green]   (\horizstart+2*\horizspace,\vertstart+0*\vertspace) rectangle (\horizstart+3*\horizspace,\vertstart+8*\vertspace);
		\draw[black,fill=green]    (\horizstart+3*\horizspace,\vertstart+0*\vertspace) rectangle (\horizstart+4*\horizspace,\vertstart+8*\vertspace);
		\draw[black,fill=yellow]   (\horizstart+4*\horizspace,\vertstart+0*\vertspace) rectangle (\horizstart+5*\horizspace,\vertstart+8*\vertspace);
		\draw[black,fill=yellow]    (\horizstart+5*\horizspace,\vertstart+0*\vertspace) rectangle (\horizstart+6*\horizspace,\vertstart+8*\vertspace);
		\draw[black,fill=yellow] (\horizstart+6*\horizspace,\vertstart+0*\vertspace) rectangle (\horizstart+7*\horizspace,\vertstart+8*\vertspace);
		\draw[black,fill=yellow]     (\horizstart+7*\horizspace,\vertstart+0*\vertspace) rectangle (\horizstart+8*\horizspace,\vertstart+8*\vertspace);
		\draw[black,fill=white]   (\horizstart+8*\horizspace,\vertstart+0*\vertspace) rectangle (\horizstart+9*\horizspace,\vertstart+8*\vertspace);
		\draw[black,fill=white]  (\horizstart+9*\horizspace,\vertstart+0*\vertspace) rectangle (\horizstart+10*\horizspace,\vertstart+8*\vertspace);
		\draw[black,fill=white]   (\horizstart+10*\horizspace,\vertstart+0*\vertspace) rectangle (\horizstart+11*\horizspace,\vertstart+8*\vertspace);
		\draw[black,fill=white]    (\horizstart+11*\horizspace,\vertstart+0*\vertspace) rectangle (\horizstart+12*\horizspace,\vertstart+8*\vertspace);
		\draw[black,fill=blue]   (\horizstart+12*\horizspace,\vertstart+0*\vertspace) rectangle (\horizstart+13*\horizspace,\vertstart+8*\vertspace);
		\draw[black,fill=blue]    (\horizstart+13*\horizspace,\vertstart+0*\vertspace) rectangle (\horizstart+14*\horizspace,\vertstart+8*\vertspace);
		\draw[black,fill=blue] (\horizstart+14*\horizspace,\vertstart+0*\vertspace) rectangle (\horizstart+15*\horizspace,\vertstart+8*\vertspace);
		\draw[black,fill=blue]     (\horizstart+15*\horizspace,\vertstart+0*\vertspace) rectangle (\horizstart+16*\horizspace,\vertstart+8*\vertspace);
		\draw[black,fill=black]   (\horizstart+16*\horizspace,\vertstart+0*\vertspace) rectangle (\horizstart+17*\horizspace,\vertstart+8*\vertspace);
		\draw[black,fill=black]  (\horizstart+17*\horizspace,\vertstart+0*\vertspace) rectangle (\horizstart+18*\horizspace,\vertstart+8*\vertspace);
		\draw[black,fill=black]   (\horizstart+18*\horizspace,\vertstart+0*\vertspace) rectangle (\horizstart+19*\horizspace,\vertstart+8*\vertspace);
		\draw[black,fill=black]    (\horizstart+19*\horizspace,\vertstart+0*\vertspace) rectangle (\horizstart+20*\horizspace,\vertstart+8*\vertspace);
		\draw[black,fill=cyan]   (\horizstart+20*\horizspace,\vertstart+0*\vertspace) rectangle (\horizstart+21*\horizspace,\vertstart+8*\vertspace);
		\draw[black,fill=cyan]    (\horizstart+21*\horizspace,\vertstart+0*\vertspace) rectangle (\horizstart+22*\horizspace,\vertstart+8*\vertspace);
		\draw[black,fill=cyan] (\horizstart+22*\horizspace,\vertstart+0*\vertspace) rectangle (\horizstart+23*\horizspace,\vertstart+8*\vertspace);
		\draw[black,fill=cyan]     (\horizstart+23*\horizspace,\vertstart+0*\vertspace) rectangle (\horizstart+24*\horizspace,\vertstart+8*\vertspace);
		\draw[black,fill=magenta]   (\horizstart+24*\horizspace,\vertstart+0*\vertspace) rectangle (\horizstart+25*\horizspace,\vertstart+8*\vertspace);
		\draw[black,fill=magenta]  (\horizstart+25*\horizspace,\vertstart+0*\vertspace) rectangle (\horizstart+26*\horizspace,\vertstart+8*\vertspace);
		\draw[black,fill=magenta]   (\horizstart+26*\horizspace,\vertstart+0*\vertspace) rectangle (\horizstart+27*\horizspace,\vertstart+8*\vertspace);
		\draw[black,fill=magenta]    (\horizstart+27*\horizspace,\vertstart+0*\vertspace) rectangle (\horizstart+28*\horizspace,\vertstart+8*\vertspace);
		\draw[black,fill=red]   (\horizstart+28*\horizspace,\vertstart+0*\vertspace) rectangle (\horizstart+29*\horizspace,\vertstart+8*\vertspace);
		\draw[black,fill=red]    (\horizstart+29*\horizspace,\vertstart+0*\vertspace) rectangle (\horizstart+30*\horizspace,\vertstart+8*\vertspace);
		\draw[black,fill=red] (\horizstart+30*\horizspace,\vertstart+0*\vertspace) rectangle (\horizstart+31*\horizspace,\vertstart+8*\vertspace);
		\draw[black,fill=red]     (\horizstart+31*\horizspace,\vertstart+0*\vertspace) rectangle (\horizstart+32*\horizspace,\vertstart+8*\vertspace);
		\draw[dashed] (\horizstart+8*\horizspace,\vertstart+8.5*\vertspace) -> (\horizstart+8*\horizspace,\vertstart-0.5*\vertspace);
		\draw[dashed] (\horizstart+16*\horizspace,\vertstart+8.5*\vertspace) -> (\horizstart+16*\horizspace,\vertstart-0.5*\vertspace);
		\draw[dashed] (\horizstart+24*\horizspace,\vertstart+8.5*\vertspace) -> (\horizstart+24*\horizspace,\vertstart-0.5*\vertspace);
		\end{tikzpicture}
	\end{center}
	\caption{The columns resulting from each row are gathered and placed in ascending order of rows.}
	\label{fig:long-shearsort-3}
\end{figure*}

When analyzing the runtime of transforming a $w \times hw$ matrix one might
think that running $h$ warps on the same streaming multiprocessor with only $w$
physical cores would increase the runtime by a factor of $h$. However, in
practice this is not the case. The reason is that as long as all data required
by the warps fits in the shared memory the context switch across the warps
incurs no overhead and running a single warp takes the same time as running $h$ 
warps with extra time spent waiting for data access due to the latency of
accessing shared memory.  

Since there are no bank conflicts in this procedure, the runtime of transforming 
an $w \times hw$ matrix remains $\OhOf{w}$ time.



%% file: pseudocode.tex
\section{Pseudocode of BCFMergesort}
\label{sec:pseudocode}
\myparagraph{Sorting $w^2$ elements in shared memory without bank conflicts.}
ShearSort sorts a matrix of values in snake-like order by first sorting the rows 
in alternating directions and then columns in ascending order. We use the fact 
that transforming a square matrix in shared memory from column-major to row-major 
order can be done by simply transposing the matrix (see Algorithm~\ref{alg:block-sort}). 
That way our implementation can assign one memory bank to each of the $w$ threads 
(see Section~\ref{sec:basecase}). For simplicity we assume that the result should 
be sorted in ascending order. Since all threads execute the same program in parallel, 
this pseudocode is written for one thread with index $tid$.

\begin{algorithm}
\caption{SHEARSORT: Bank-conflict free sorting in shared memory}
\label{alg:block-sort}
\begin{tabbing}
IN\=PU\=T:\=  \ $w\times w$ matrix $m$ \\
OUTPUT: Sorted matrix $m$ \\
\\
register array $R$[$w$] \\
FOR $i$ IN $1$\dots $\log_2(w)$ DO \+ \\
	\# Sort the rows of matrix $m$ in alternating order \\
	FOR $j$ in $0\dots (w-1)$ DO $R[j] = m[tid][j]$ \\ \\
	BATCHERS-ODD-EVEN-MERGESORT($R$) \\ \\
	IF $tid \mod 2 == 0$ THEN \+ \\
		FOR $j$ in $0\dots (w-1)$ DO $m[tid][j] = R[j]$ \- \\
	ELSE \+ \\
		FOR $j$ in $0\dots (w-1)$ DO $m[tid][w - j] = R[j]$ \- \\ \\
	
	\# Change matrix from column-major order to row-major order \\
	TRANSPOSE($m$) \\ \\
	
	\# Sort the columns of matrix $m$ \\
	FOR $j$ in $0\dots (w-1)$ DO $R[j] = m[tid][j]$ \\ \\
	BATCHERS-ODD-EVEN-MERGESORT($R$) \\ \\
	FOR $j$ in $0\dots (w-1)$ DO $m[tid][j] = R[j]$ \\ \\
	
	\# Change matrix from row-major order to column-major order \\
	TRANSPOSE($m$) \- \\
\\
\# Clean up and remove snake-like order \\
FOR $j$ in $0\dots (w-1)$ DO $R[j] = m[tid][j]$ \\ \\
BATCHERS-ODD-EVEN-MERGESORT($R$) \\ \\
FOR $j$ in $0\dots (w-1)$ DO $m[tid][j] = R[j]$ \\ \\

\# Change matrix from column-major order to row-major order \\
TRANSPOSE($m$)
\end{tabbing}
\end{algorithm}

As explained in Section~\ref{sec:optimizations}, the transpositions can be removed 
if we treat diagonals as columns and shifted columns as rows. The pseudocode for 
this optimization can be seen in Algorithm~\ref{alg:diagonal-sort}. The extra step 
of transforming diagonals to rows at the end of the algorithm is easily amortized
by the missing $2\cdot\log_2(w) + 1$ transpositions.

\begin{algorithm}
\caption{SHEARSORT-DIAGONAL: Bank-conflict free sorting in shared memory without transpositions}
\label{alg:diagonal-sort}
\begin{tabbing}
IN\=PU\=T:\=  \ $w\times w$ matrix $m$ \\
OUTPUT: Sorted matrix $m$ \\
\\
register array $R$[$w$] \\
FOR $i$ IN $1$\dots $\log_2(w)$ DO \+ \\
	\# Sort the diagonals of matrix $m$ in alternating order \\
	FOR $j$ in $0$\dots $(w-1)$ DO $R$[$j$] = $m$[$(tid + j) \mod w$][$j$] \\ \\
	BATCHERS-ODD-EVEN-MERGESORT($R$) \\ \\
	IF $tid \mod 2 == 0$ THEN \+ \\
		FOR $j$ in $0$\dots $(w-1)$ DO $m$[$(tid + j) \mod w$][$j$] = $R$[$j$] \- \\
	ELSE \+ \\
		FOR $j$ in $0$\dots $(w-1)$ DO $m$[$(tid + w - j) \mod w$][$w - j$] = $R$[$j$] \- \\ \\
	
	\# Sort the shifted columns of matrix $m$ \\
	FOR $j$ in $0$\dots $(w-1)$ DO $R$[$j$] = $m$[$(tid + j) \mod w$][$tid$] \\ \\
	BATCHERS-ODD-EVEN-MERGESORT($R$) \\ \\
	FOR $j$ in $0$\dots $(w-1)$ DO $m$[$(tid + j) \mod w$][$tid$] = $R$[$j$] \- \\
\\
\# Remove snake-like order \\
FOR $j$ in $0$\dots $(w-1)$ DO $R$[$j$] = $m$[$(tid + j) \mod w$][$j$] \\ \\
BATCHERS-ODD-EVEN-MERGESORT($R$) \\ \\
FOR $j$ in $0$\dots $(w-1)$ DO $m$[$(tid + j) \mod w$][$j$] = $R$[$j$] \\ \\

\# Transform diagonals to columns \\
FOR $j$ in $0$\dots $(w-1)$ DO $R$[$j$] = $m$[$(2 \cdot tid + j) \mod w$][$(j + tid) \mod w$] \\
FOR $j$ in $0$\dots $(w-1)$ DO $m$[$(tid + j) \mod w$][$tid$] = $R$[$j$]
\end{tabbing}
\end{algorithm}

\myparagraph{Sorting $h\cdot w^2$ elements in shared memory without bank conflicts.}
We extend our ShearSort implementation to matrices of size $w \times (h \cdot w)$. 
We define the functions {\tt CHANGE-TO-ROWMAJOR} 
(Algorithm~\ref{alg:change-to-rowmajor}) and {\tt CHANGE-TO-COLUMNMAJOR} 
(Algorithm~\ref{alg:change-to-columnmajor}) that transforms the long rows into 
$h$ columns of size $w$. These functions require $h$ warps.

\begin{algorithm}
\caption{CHANGE-TO-ROWMAJOR: Turn column-major long matrices to row-major order}
\label{alg:change-to-rowmajor}
\begin{tabbing}
IN\=PU\=T:\=  \ $w\times (wh)$ matrix $m$ \\
OUTPUT: Transformed matrix $m$ \\
\\
$widx = \lfloor\frac{tid}{w}\rfloor$ \\
$wpos = tid\mod w$ \\ \\

\# Change all $w\times w$ submatrices $m_1\dots m_h$ of matrix $m$ to row-major order \\
TRANSPOSE($m_{widx}$) \\ \\

register array $R$[$w$] \\

FOR $i$ in $0\dots \left(\frac{w}{h}-1\right)$ DO \+ \\
	FOR $j$ in $0\dots (h-1)$ DO \+ \\
		$R$[$h \cdot i + j$] = $m$[$wpos$][$(widx \cdot \frac{w}{h} + i) + j \cdot w$] \- \- \\ \\

FOR $i$ in $0\dots (w-1)$ DO \+ \\
	$m$[$wpos$][$widx \cdot w + i$] = $R$[$i$]
\end{tabbing}
\end{algorithm}

\begin{algorithm}
\caption{CHANGE-TO-COLUMNMAJOR: Turn row-major long matrices to column-major order}
\label{alg:change-to-columnmajor}
\begin{tabbing}
IN\=PU\=T:\=  \ $w\times (wh)$ matrix $m$ \\
OUTPUT: Transformed matrix $m$ \\
\\
$widx = \lfloor\frac{tid}{w}\rfloor$ \\
$wpos = tid \mod w$ \\ \\

register array $R$[$w$] \\ 

FOR $i$ in $0\dots (w-1)$ DO \+ \\
	$R$[$i$] = $m$[$wpos$][$widx \cdot w + i$] \- \\ \\

FOR $i$ in $0\dots \left(\frac{w}{h}-1\right)$ DO \+ \\
	FOR $j$ in $0\dots (h-1)$ DO \+ \\
		$m$[$wpos$][$(widx \cdot \frac{w}{h} + i) + j \cdot w$] = $R$[$h \cdot i + j$] \- \- \\ \\

\# Change all $w\times w$ submatrices $m_1\dots m_h$ of matrix $m$ to row-major order \\
TRANSPOSE($m_{widx}$)

\end{tabbing}
\end{algorithm}

Furthermore, {\tt SHEARSORT} has to be modified to handle the long rows 
(Algorithm~\ref{alg:block-sort-long}). For this procedure to run, we need a 
subroutine {\tt SHEARSORT-SEGMENTED} that takes three parameters: the matrix, 
the number of rows per segment and a flag snake-like that indicates whether the 
segments should be rearranged in the snake-like order required by ShearSort.
{\tt SHEARSORT-SEGMENTED} performs {\tt SHEARSORT} where the second sort (the 
columns) is replaced by a {\em segmented sort} routine that sorts $\frac{w}{h}$ 
segments of size $h$. 

For the diagonal version of BCFMergesort, changing the matrix to a different 
memory representation is reduced to gathering and scattering the diagonals in the
same manner as {\tt CHANGE-TO-ROWMAJOR} or {\tt CHANGE-TO-COLUMNMAJOR} treat the columns.

\begin{algorithm}
\caption{SHEARSORT-LONG: Bank-conflict free sorting of long matrices in shared memory}
\label{alg:block-sort-long}
\begin{tabbing}
IN\=PU\=T:\=  \ $w\times (wh)$ matrix $m$ \\
OUTPUT: Sorted matrix $m$ \\
\\
register array $R$[$w$] \\
FOR $i$ IN $1$\dots $\log_2(w)$ DO \+ \\
	\# Change matrix to row-major order \\
	CHANGE-TO-ROWMAJOR($m$) \\ \\
	
	\# Change all segmented $w\times w$ submatrices $m^s_1\dots m^s_h$ from row-major order to column-major order \\
	TRANSPOSE($m^s_{widx}$) \\ \\
	
	\# Segmented sort submatrices $m^s_1\dots m^s_h$ of matrix $m$ in snake-like order \\
	SHEARSORT-SEGMENTED($m_{widx}$, $h$, true) \\ \\
	
	\# Change matrix to column-major order \\
	CHANGE-TO-COLUMNMAJOR($m$) \\ \\
	
	\# Change submatrices $m_1\dots m_h$ from column-major order to row-major order \\
	TRANSPOSE($m_{widx}$) \\ \\
	\# Sort the columns of all submatrices $m_1\dots m_h$ \\
	FOR $j$ in $0$\dots $(w-1)$ DO $R$[$j$] = $m_{widx}$[$tid$][$j$] \\ 
	BATCHERS-ODD-EVEN-MERGESORT($R$) \\ 
	FOR $j$ in $0$\dots $(w-1)$ DO $m_{widx}$[$tid$][$j$] = $R$[$j$] \\ \\
	
	\# Change submatrices $m_1\dots m_h$ from row-major order to column-major order \\
	TRANSPOSE($m_{widx}$) \- \\ \\

\# Change matrix to row-major order \\
CHANGE-TO-ROWMAJOR($m$) \\ \\

\# Change segmented submatrices $m^s_1\dots m^s_h$ from row-major order to column-major order \\
TRANSPOSE($m^s_{widx}$) \\ \\

\# Segmented sort submatrices $m^s_1\dots m^s_h$ in ascending order \\
SHEARSORT-SEGMENTED($m_{widx}$, $h$, false) 
\end{tabbing}
\end{algorithm}

\myparagraph{Page-wise binary merge.}
To perform the bank conflict free page-wise binary merge (see Section~\ref{sec:merging}),
we load one page of data from both input streams, merge them and output one full page. 
Figuratively, the keys from the first exhausted page are moved to the other stream and 
are replaced keys from that stream. That is why merging has to continue until both 
streams are exhausted and cannot be stopped when the first input stream is finished. 

For a better performance we modify the {\tt SHEARSORT} subroutine to only perform
one round of ShearSort instead of $\log_2(w)$. By the 0-1 principle this is enough 
to merge two sorted sequences. We call this shortened sorting routine SHEARSORT-MERGE($S$).

In the case of arbitrary input stream sizes, we have to determine
the amount of data to load (e.g. by setting an integer $ls = \min\left(\frac{M}{2}, n[e]-ic[e]\right)$)
and in case of partially filled pages pad them with $\infty$. Since the padding will 
remain in shared memory until the merge is finished, we also have to keep track 
of the amount of padding used and subtract it from the size of the final outputs.

\begin{algorithm}
\caption{BINARY-MERGE: Page-wise binary merge}
\label{alg:binary-merge}
\begin{tabbing}
IN\=PU\=T:\=  \ Streams $I[0]$ and $I[1]$, stream size $n$ \\
OUTPUT: Output stream $O$ \\
\\
shared memory array $S$[$M$] \\
integer e \\ \\

\# Load a page of $\frac{M}{2}$ keys from each input stream into $S$ \\
$S[0\dots \left(\frac{M}{2}-1\right)] = I[0][0\dots \left(\frac{M}{2}-1\right)]$ \\
$S[\frac{M}{2}\dots (M-1)] = I[1][0\dots \left(\frac{M}{2}-1\right)]$ \\ \\

\# Determine the input stream, whose page will be exhausted first \\
IF $S[\frac{M}{2}-1] \leq S[M-1]$ THEN \+ \\
	e = 0 \- \\
ELSE \+ \\
	e = 1 \- \\ \\

\# Merge the first pages of both input streams \\
SHEARSORT-MERGE($S$) \\ \\

\# Write the first page to the output stream \\
$O[0\dots \left(\frac{M}{2}-1\right)] = S[0\dots \left(\frac{M}{2}-1\right)]$ \\ \\

\# Offset counters for all streams \\
integer $ic[0] = ic[1] = oc = \frac{M}{2}$ \\ \\

WHILE $(ic[0] < n)$ OR $(ic[1] < n)$ DO \+ \\
	\# If there are elements left in the designated input stream, load a page. \\
	\# Otherwise load a page from the other stream. Adjust the offset counter.\\
	IF $ic[e] \geq n$ THEN $e = 1 - e$ \\
	$S[0\dots \left(\frac{M}{2}-1\right)] = I[e][ic[e]\dots \left(ic[e]+\frac{M}{2}-1\right)]$ \\
	$ic[e] = ic[e] + \frac{M}{2}$ \\ \\
	
	\# Determine which is the next page to be exhausted \\
	IF $S[\frac{M}{2}-1] > S[M-1]$ THEN $e = 1 - e$ \\ \\
	
	\# Merge the pages in shared memory \\
	SHEARSORT-MERGE($S$) \\ \\

	\# Write the next page to the output stream and adjust the offset counter \\
	$O[oc\dots \left(oc+\frac{M}{2}-1\right)] = S[0\dots \left(\frac{M}{2}-1\right)]$ \\
	$oc = oc+\frac{M}{2}$ \- \\ \\
	
\# Write the final remaining page to the output stream \\
$O[oc\dots (2\cdot n-1)] = S[\frac{M}{2}\dots (M-1)]$

\end{tabbing}
\end{algorithm}

\myparagraph{Bucket distribution.}
The bucket distribution of Section~\ref{sec:merging} consists of two phases: finding 
the splitters and determining their positions in the input data, and rearranging the 
data. The second part is not strictly necessary but it makes life easier when the 
merging begins again.

The subroutine {\tt FIND-BUCKETS} is called on $P$ CTAs, identified by their 
index $pid$, with $w$ threads each. The sorted substreams of input stream $I$ are 
marked as $I_1\dots I_P$. We also need a function {\tt EXCLUSIVE-PREFIX-SUMS} that 
performs the namesake operation on $P$ integers. Since $P = 128$ in our experiments,
we chose a simple linear solution executed on one thread.

\begin{algorithm}
\caption{FIND-BUCKETS: Bucket distribution}
\label{alg:find-buckets}
\begin{tabbing}
IN\=PU\=T:\=  \ Input stream $I$, consisting of $P$ sorted substreams of size $n$, number of pivots $t$ \\
OUTPUT: Output stream $O$, stream separators $Q$ \\
\\
global potential pivot array $S[P\cdot t]$ \\
global final pivot array $F[P]$ \\
global temporary splitter array $T[P^2]$ \\
global bucket offset array $B[P]$ \\
integer $c$, $oc$ \\ \\

\# Pick $t$ potential splitters per substream $I_k$ \\
FOR $i$ in $0\dots t$ DO $S[pid\cdot t+i] = I_{pid}[i\cdot\frac{n}{t}]$ \\ \\

\# Sort all potential pivots \\
MERGESORT($S$) \\ \\

\# Pick the final pivots: Every $t$th element of $S$ \\
$F[pid] = A[pid \cdot t]$ \\ \\

\# Determine the rank of each pivot in each substream \\
$c = 1$ \\
FOR $i$ in $0\dots (n-1)$ DO \+ \\
	IF $I_{pid}[i] > F[c]$ THEN \+ \\
		$T[pid \cdot P + ] = pid \cdot n + i$ \\
		WHILE $I_{pid}[i] > F[c]$ DO $c = c + 1$ \- \- \\ \\

\# Calculate target substream offsets (sum entries of each bucket, run prefix sums) \\
$B[pid] = (T[0\cdot P+pid+1]-T[0\cdot P+pid]) + \dots + (T[(P-1)\cdot P+pid+1]-T[(P-1)\cdot P+pid])$ \\ \\

EXCLUSIVE-PREFIX-SUMS($B$) \\ \\

\# Sort substreams by bucket \\
$oc = B[pid]$ \\
FOR $i$ in $0\dots (P-1)$ DO \+ \\
	$O[oc\dots (oc+(T[pid + i\cdot P + 1]-T[pid + i\cdot P]))] = $ \+ \+ \\
			$I[T[pid + i\cdot P]]\dots I[T[pid + i\cdot P + 1]-1]$ \- \- \\
	$Q[pid \cdot P + i] = oc$ \\
	$oc = (oc+(T[pid + i\cdot P + 1]-T[pid + i\cdot P]))$

\end{tabbing}
\end{algorithm}

%% file: esa2014.bbl
\begin{thebibliography}{10}
\providecommand{\url}[1]{\texttt{#1}}
\providecommand{\urlprefix}{URL }

\bibitem{aggarwal:io-model}
Aggarwal, A., Vitter, J.: {The Input/Output Complexity of Sorting and Related
  Problems}. Communications of the ACM pp. 1116--1127 (1988)

\bibitem{batcher:sorting}
Batcher, K.E.: {Sorting Networks and their Applications}. In: Proceedings of
  the AFIPS Spring Joint Computer Conference. pp. 307--314. ACM (1968)

\bibitem{bell:thrust}
Bell, N., Hoberock, J.: {Thrust: A productivity-oriented library for CUDA}. GPU
  Computing Gems: Jade Edition pp. 359--372 (2011)

\bibitem{cederman2008practical}
Cederman, D., Tsigas, P.: {A Practical Quicksort Algorithm for Graphics
  Processors}. In: Algorithms-ESA 2008, pp. 246--258. Springer (2008)

\bibitem{davidson2012efficient}
Davidson, A., Tarjan, D., Garland, M., Owens, J.D.: {Efficient Parallel Merge
  Sort for Fixed and Variable Length Keys}. In: Innovative Parallel Computing
  (InPar), 2012. pp. 1--9. IEEE (2012)

\bibitem{dotsenko:scan}
Dotsenko, Y., Govindaraju, N.K., Sloan, P.P., Boyd, C., Manfedelli, J.: {Fast
  Scan Algorithms on Graphics Processors}. In: Proceedings of the 22nd Annual
  International Conference on Supercomputing (2008)

\bibitem{knuth:seminumerical}
Knuth, D.E.: Seminumerical Algorithms, The Art of Computer Programming, vol.~2.
  Addison-Wesley, Reading, third edn. (1998)

\bibitem{knuth:sorting}
Knuth, D.E.: Sorting and Searching, The Art of Computer Programming, vol.~3.
  Addison-Wesley, Reading, third edn. (1998)

\bibitem{gpu-sample-sort}
Leischner, N., Osipov, V., Sanders, P.: {GPU Sample Sort}. In: Proceedings of
  the 24th IEEE Parallel and Distributed Processing Symposium (IPDPS) (2010)

\bibitem{merrill:scan}
Merrill, D., Grimshaw, A.: {Parallel Scan for Stream Architectures}. Tech. Rep.
  CS2009-14, Department of Computer Science, University of Virginia (2009)

\bibitem{merrill2011high}
Merrill, D., Grimshaw, A.: {High Performance and Scalable Radix Sorting: A Case
  Study of Implementing Dynamic Parallelism for GPU Computing}. Parallel
  Processing Letters  21(02),  245--272 (2011)

\bibitem{cudaprogrammingguide}
{NVIDIA}: {NVIDIA CUDA Programming Guide 5.0} (2012),
  \url{http://docs.nvidia.com/cuda/pdf/CUDA_C_Programming_Guide.pdf}, last
  viewed on 2/11/2013

\bibitem{peters2010fast}
Peters, H., Schulz-Hildebrandt, O., Luttenberger, N.: {Fast In-Place Sorting
  with CUDA based on Bitonic Sort}. In: Parallel Processing and Applied
  Mathematics, pp. 403--410. Springer (2010)

\bibitem{purcell2003photon}
Purcell, T.J., Donner, C., Cammarano, M., Jensen, H.W., Hanrahan, P.: {Photon
  Mapping on Programmable Graphics Hardware}. In: Proceedings of the ACM
  SIGGRAPH/EUROGRAPHICS conference on Graphics hardware. pp. 41--50.
  Eurographics Association (2003)

\bibitem{satish:mergesort}
Satish, N., Harris, M., Garland, M.: {Designing Efficient Sorting Algorithms
  for Manycore GPUs}. In: Proceedings of the 23rd IEEE Parallel and Distributed
  Processing Symposium (IPDPS). pp. 1--10. IEEE (2009)

\bibitem{scherson:shear-sort}
Sen, S., Scherson, I.D., Shamir, A.: {Shear Sort: A True Two-Dimensional
  Sorting Techniques for VLSI Networks}. In: ICPP. pp. 903--908 (1986)

\bibitem{sintorn2008fast}
Sintorn, E., Assarsson, U.: {Fast Parallel GPU-Sorting using a Hybrid
  Algorithm}. Journal of Parallel and Distributed Computing  68(10),
  1381--1388 (2008)

\bibitem{ye:sort}
Ye, X., Fan, D., Lin, W., Yuan, N., Ienne, P.: {High Performance
  Comparison-Based Sorting Algorithm on Many-Core GPUs}. In: Proceedings of the
  24th IEEE Parallel and Distributed Processing Symposium (IPDPS) (2010)

\end{thebibliography}
